\documentclass[twocolumn,showpacs,preprintnumbers,amsmath,amssymb,floatfix,prc]{revtex4}
\usepackage{graphicx}
\usepackage{dcolumn}
\usepackage{bm}
\usepackage{float}
\begin{document}
\preprint{APS/123-QED}
\title{Mass measurements in the vicinity of the doubly-magic waiting point $^{56}$Ni}
\author{A.~Kankainen} \email{anu.k.kankainen@jyu.fi}  
\affiliation{Department of Physics, University of Jyv\"askyl\"a, P.O. Box 35, FI-40014 University of Jyv\"askyl\"a, Finland}
\author{V.-V.~Elomaa}\altaffiliation[Present address:]{~Turku PET Centre, Accelerator Laboratory, \AA bo Akademi University, FI-20500 Turku, Finland}
\affiliation{Department of Physics, University of Jyv\"askyl\"a, P.O. Box 35, FI-40014 University of Jyv\"askyl\"a, Finland}
\author{T.~Eronen}
\affiliation{Department of Physics, University of Jyv\"askyl\"a, P.O. Box 35, FI-40014 University of Jyv\"askyl\"a, Finland}
\author{D.~Gorelov}
\affiliation{Department of Physics, University of Jyv\"askyl\"a, P.O. Box 35, FI-40014 University of Jyv\"askyl\"a, Finland}
\author{J.~Hakala}
\affiliation{Department of Physics, University of Jyv\"askyl\"a, P.O. Box 35, FI-40014 University of Jyv\"askyl\"a, Finland}
\author{A.~Jokinen}
\affiliation{Department of Physics, University of Jyv\"askyl\"a, P.O. Box 35, FI-40014 University of Jyv\"askyl\"a, Finland}
\author{T.~Kessler}
\affiliation{Department of Physics, University of Jyv\"askyl\"a, P.O. Box 35, FI-40014 University of Jyv\"askyl\"a, Finland}
\author{V.S.~Kolhinen}
\affiliation{Department of Physics, University of Jyv\"askyl\"a, P.O. Box 35, FI-40014 University of Jyv\"askyl\"a, Finland}
\author{I.D.~Moore}
\affiliation{Department of Physics, University of Jyv\"askyl\"a, P.O. Box 35, FI-40014 University of Jyv\"askyl\"a, Finland}
\author{S.~Rahaman} \altaffiliation[Present address:]{~Physics Division, P-23, Mail Stop H803, Los
Alamos National Laboratory, Los Alamos, NM 87545, USA.}
\affiliation{Department of Physics, University of Jyv\"askyl\"a, P.O. Box 35, FI-40014 University of Jyv\"askyl\"a, Finland}
\author{M.~Reponen}
\affiliation{Department of Physics, University of Jyv\"askyl\"a, P.O. Box 35, FI-40014 University of Jyv\"askyl\"a, Finland}
\author{J.~Rissanen}
\affiliation{Department of Physics, University of Jyv\"askyl\"a, P.O. Box 35, FI-40014 University of Jyv\"askyl\"a, Finland}
\author{A.~Saastamoinen}
\affiliation{Department of Physics, University of Jyv\"askyl\"a, P.O. Box 35, FI-40014 University of Jyv\"askyl\"a, Finland}
\author{C.~Weber} \altaffiliation[Present address:]{~Department of Physics, Ludwig-Maximilians-Universit\"at M\"unchen, D-85748 Garching, Germany}
\affiliation{Department of Physics, University of Jyv\"askyl\"a, P.O. Box 35, FI-40014 University of Jyv\"askyl\"a, Finland}
\author{J.~\"Ayst\"o}
\affiliation{Department of Physics, University of Jyv\"askyl\"a, P.O. Box 35, FI-40014 University of Jyv\"askyl\"a, Finland}

\date{\today}

\begin{abstract}
Masses of $^{56,57}$Fe, $^{53}$Co$^m$, $^{53,56}$Co, $^{55,56,57}$Ni, $^{57,58}$Cu, and $^{59,60}$Zn have been determined with the JYFLTRAP Penning trap mass spectrometer at IGISOL with a precision of $\delta m/m\leq 3\times10^{-8}$. The $Q_\text{EC}$ values for $^{53}$Co, $^{55}$Ni, $^{56}$Ni, $^{57}$Cu, $^{58}$Cu, and $^{59}$Zn have been measured directly with a typical precision of better than $0.7~\text{keV}$ and Coulomb displacement energies have been determined. The $Q$ values for proton captures on $^{55}$Co, $^{56}$Ni, $^{58}$Cu, and $^{59}$Cu have been measured directly. The precision of the proton-capture $Q$ value for $^{56}\text{Ni}(p,\gamma)^{57}\text{Cu}$, $Q_{(p,\gamma)}=689.69(51)~\text{keV}$, crucial for astrophysical $rp$-process calculations, has been improved by a factor of 37. The excitation energy of the proton-emitting spin-gap isomer $^{53}$Co$^m$ has been measured precisely, $E_x=3174.3(10)~\text{keV}$, and a Coulomb energy difference of $133.9(10)~\text{keV}$ for the $19/2^-$ state has been obtained. Except for $^{53}$Co, the mass values have been adjusted within a network of 17 frequency ratio measurements between 13 nuclides which allowed also a determination of the reference masses $^{55}$Co, $^{58}$Ni, and $^{59}$Cu.
\end{abstract}


\maketitle

\section{\label{sec:Intro}Introduction}

$^{56}$Ni is a waiting-point nucleus in the astrophysical rapid-proton capture process ($rp$ process) which occurs at high temperatures and high hydrogen densities (see \emph{e.g.}, Ref.~\cite{Sch98}). In the $rp$ process, nuclides capture protons until they are inhibited by a low or negative $Q$ value. At such points, the process must proceed via much slower beta decay. At $^{56}$Ni, the proton-capture $Q$ value to  $^{57}$Cu is quite low and critical for the synthesis of elements heavier than nickel. Namely, the beta-decay half-life of $^{56}$Ni is 6.075(10) days \cite{Cru92}, exceeding all normal time scales of x-ray bursts and other places where the $rp$ process could occur. Previously, $^{56}$Ni was considered as the end point of the $rp$ process \cite{Wal81}, but later it was shown to proceed until the SnSbTe-region \cite{Sch01,Elo09a}. For an accurate modeling of this process, the proton-capture $Q$ value for the reaction $^{56}$Ni(p,$\gamma$)$^{57}$Cu has to be known precisely.

$^{56}$Ni is doubly-magic and therefore, the precise knowledge of its mass and the masses of the neighboring nuclei is important for nuclear structure studies around $Z=N=28$. Nuclei close to or at the $N=Z$ line offer an interesting possibility to study the exchange symmetry between neutrons and protons. The $Q_\text{EC}$ values between the isospin $T=1/2$ mirror nuclei provide direct information on the Coulomb displacement energies (CDE), in other words, the binding energy differences between two adjacent members of an isobaric multiplet. By plotting the Coulomb energy differences (CED), \emph{i.e.} the differences in the level excitation energies of mirror nuclei, as a function of the spin, interesting information on changes in nuclear structure can be obtained. One of the mirror nuclei close to $^{56}$Ni is $^{53}$Co which has a renowned spin-gap isomer $^{53}$Co$^m$ $(19/2^-)$ from which direct proton decay was observed for the first time \cite{Cer70,Cer72}. A precise and direct measurement of this excitation energy is needed for an accurate Coulomb energy difference value of the $^{53}$Co $19/2^-$ state. 

Recently, the $Q_\text{EC}$ values of lighter $T=1/2$ nuclei have been used to determine high-precision corrected $ft$ values. From the corrected $ft$ values, a mixing ratio of Fermi and Gamow-Teller transitions is obtained \cite{Sev08}. This mixing ratio is useful for testing the Standard Model values for the beta-decay correlation coefficients \cite{Sev08}, such as the beta-neutrino angular correlation coefficient. If the beta asymmetry parameter $A_\beta$, neutrino asymmetry parameter $B_\nu$, or beta-neutrino angular correlation coefficient $a_{\beta\nu}$ has already been measured, the mixing ratio can be determined and the $|V_{ud}|$ value for the Cabibbo-Kobayashi-Maskawa (CKM) matrix can be extracted from the corrected $ft$ values \cite{Nav08}. This, in turn, provides an opportunity to test the conserved vector current (CVC) hypothesis.

\section{\label{sec:exp}Experimental methods}
The studied neutron-deficient nuclides were produced at the Ion-Guide Isotope Separator On-Line (IGISOL) facility
\cite{Ays01}. In the first run, proton or $^3$He$^{2+}$ beams from the K-130 cyclotron impinging
on enriched $^{54}$Fe ($2~\text{mg/cm}^2$) or $^{58}$Ni ($1.8~\text{mg/cm}^2$) targets produced the ions of interest employing the 
light-ion ion-guide \cite{Elo08}. The corresponding proton beam intensity was about $10~\mu A$ and the $^3$He$^{2+}$ beam $0.5~p\mu A$. A 50 MeV 
proton beam was used to test the production of $^{54}$Ni and $^{56}$Cu. However, these exotic nuclides were not observed in this run. The properties of the studied nuclides are summarized in Table~\ref{tab:properties} and the production methods in Table~\ref{tab:ratios}. 

In the second run, the ions of interest were searched for via heavy-ion fusion-evaporation reactions with a $^{20}$Ne$^{4+}$ beam impinging on a 
calcium target ($4~\text{mg/cm}^2$) at $75~\text{MeV}$ and $105~\text{MeV}$. Previously, the heavy-ion ion-guide (HIGISOL) \cite{Hui02} 
has been successfully used for producing heavier nuclides for JYFLTRAP mass measurements \cite{Kan05,Web08}. This was the first experiment 
performed in a lighter mass region. At HIGISOL, the target wheel is located along the cyclotron beam line before the gas cell and the primary heavy-ion beam is stopped in a graphite beam dump before entering the cell to avoid plasma effects. This sets two requirements for the recoiling reaction products: they have to scatter at large enough angles and they have to have sufficient energy to pass through the entrance window around the gas cell. The $^{20}$Ne$+^{40}$Ca reaction gave enough angular spread for the recoils but not enough energy for them to pass sufficiently through a $2~\text{mg/cm}^2$-thick 
Havar entrance window to the HIGISOL gas cell. Therefore, only the reference nuclides $^{57}$Ni and $^{56}$Co were measured against the reference $^{58}$Ni in this latter run. To complete the network of the measured frequency ratios, stable reference ions $^{56}$Fe$^+$, $^{57}$Fe$^+$, and $^{58}$Ni$^+$ were produced with an offline electric discharge ion source \cite{Rah08} at IGISOL and measured against each other.

\begin{table}[!]
\caption{\label{tab:properties} Properties of the nuclides studied in this work taken from Ref.~\cite{nubase}. Given are the half-lives 
($T_{1/2}$), spins ($I$), parities ($\pi$), and excitation energies of the isomers ($E_x$).}
\begin{ruledtabular}
\begin{tabular}{llll}
Nuclide & $T_{1/2}$ & $I^{\pi}$ & $E_x$ (keV)\\
\hline
$^{56}$Fe & stable & $0^+$& \\
$^{57}$Fe & stable & $1/2^-$ & \\
$^{53}$Co & 244.6(76) ms \footnotemark[1] & $7/2^-$\# & \\
$^{53}$Co$^m$ & 247(12) ms & $(19/2^-)$ & 3197(29)\\
$^{55}$Co & 17.53(3) h & $7/2^-$ & \\
$^{56}$Co & 77.23(3) d & $4^+$ & \\
$^{55}$Ni & 203.3(37) ms \footnotemark[1] & $7/2^-$ & \\
$^{56}$Ni & 6.075 (10) d & $0^+$ & \\
$^{57}$Ni & 35.60(6) h & $3/2^-$ & \\
$^{58}$Ni & stable & $0^+$& \\
$^{57}$Cu & 196.44(68) ms \footnotemark[1] & $3/2^-$ & \\
$^{58}$Cu & 3.204(7) s & $1^+$ & \\
$^{59}$Cu & 81.5(5) s & $3/2^-$ & \\
$^{59}$Zn & 181.9(18) ms \footnotemark[1] & $3/2^-$ & \\
$^{60}$Zn & 2.38(5) min & $0^+$ & \\
\end{tabular}
\footnotetext[1]{The half-life taken from Ref.~\cite{Sev08}.}
\end{ruledtabular}
\end{table}

After extraction from the gas cell, the ions were accelerated to 30 keV and mass-separated by a $55^\circ$ dipole magnet. The ions with the same mass number $A$ were sent to a radio-frequency quadrupole (RFQ) cooler and buncher \cite{Nie01} which delivered the ions as short, cooled bunches to the JYFLTRAP 
Penning trap mass spectrometer \cite{Kol04}. JYFLTRAP consists of two cylindrical Penning traps inside a $B=7~\text{T}$ superconducting solenoid. The first 
trap, the \emph{purification trap}, is used for selecting the isobar (in some cases even the isomer) of interest via mass-selective buffer gas 
cooling \cite{Sav91}. After the first trap, the ions were sent to the second trap, the \emph{precision trap}, where the masses of the ions 
$m$ with a charge $q$ were measured precisely by determining the cyclotron frequency $\nu_c=qB/(2\pi m)$ via a time-of-flight (TOF) ion cyclotron resonance method \cite{Gra80,Kon95}. The cyclotron frequency was obtained by measuring the sideband frequency $\nu_+ + \nu_-$, where $\nu_+$ and $\nu_-$ are the reduced cyclotron and magnetron frequencies, respectively. The sideband frequency corresponds to the cyclotron frequency with such a high precision that it can be used in the mass measurements \cite{Gab09}.

Conventionally, the resonance curve is obtained with a quadrupolar RF field with a typical 
duration of $200-800~\text{ms}$. Recently, a Ramsey method of time-separated oscillatory fields has been applied to short-lived ions in Penning
traps \cite{Geo07,Kre07}. The Ramsey method decreases the linewidth of the resonance and makes the sidebands much stronger resulting in a considerably smaller statistical uncertainty in the cyclotron frequency. The Ramsey excitation scheme and a new method of Ramsey cleaning have been successfully applied at JYFLTRAP \cite{Ero09}. In the new cleaning mode, the ions from the purification trap are excited by a time-separated oscillatory electric dipole field in the precision trap. The undesired ions are driven into a larger orbit while the ions of interest remain unaffected if an appropriate dipole excitation pattern is chosen. After that, the ions of interest are sent back to the purification trap whereas the unwanted ions cannot pass through the 2-mm diaphragm between the traps. In the purification trap, the ions of interest are recentered and returned once more to the precision trap for the final mass measurement. A so-called back-and-forth scheme is similar to the Ramsey cleaning scheme except that no dipole excitation in the precision trap is applied before sending the ions back to the purification trap resulting in much smaller bunch size.

In this work, normal (conventional) TOF resonances were measured for all ions in order to be sure about the center frequency in the Ramsey 
excitation scheme. The duration of the Ramsey fringes was $25~\text{ms}$. The waiting time between the two fringes was $350~\text{ms}$  for the long-lived nuclide $^{56}$Ni and its references and $150~\text{ms}$ for the other, shorter-lived nuclides and their references. Only $^{60}$Zn and its references were measured with the conventional TOF method with a quadrupolar excitation period of $800~\text{ms}$. Ramsey cleaning with a $25~\text{ms}-30~\text{ms}-25~\text{ms}$ dipole excitation scheme was applied for $^{55}$Co and $^{55}$Ni. For $^{56}$Ni, $^{57,58}$Cu and $^{59}$Zn and their references (except $^{55}$Co) a back-and-forth scheme was used. In the HIGISOL run, a $25~\text{ms}-150~\text{ms}-25~\text{ms}$ excitation pattern with the back-and-forth purification was used for $^{56}$Co, $^{57}$Ni and $^{58}$Ni. In the run employing the electric discharge ion source, a $25~\text{ms}-350~\text{ms}-25~\text{ms}$ excitation pattern with the back-and-forth purification was used for $^{56,57}$Fe, and $^{58}$Ni.

\section{\label{sec:data}Data analysis}

\subsection{\label{sec:ratio_analysis}Analysis of the measured frequency ratios}

\begin{figure}[!]
\centering
\includegraphics[width=0.45\textwidth,clip]{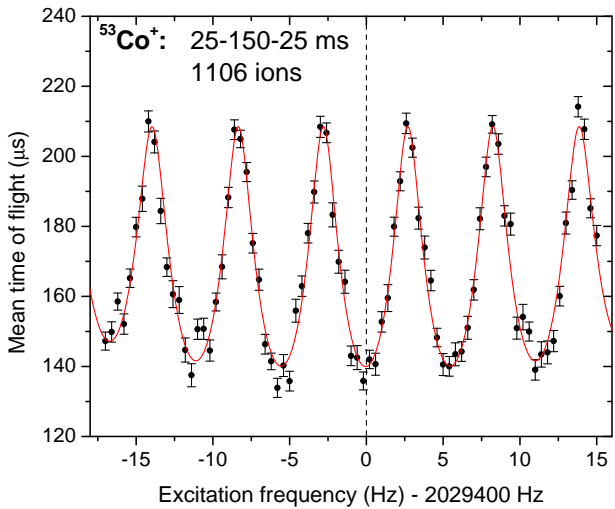}
\caption{(Color online) Cyclotron resonance curve for $^{53}\text{Co}^+$. Ramsey excitation with a $25~\text{ms}-150~\text{ms}-25~\text{ms}$ (on-off-on) pattern was used. Only bunches having one single ion are shown.}
\label{fig:TOF}
\end{figure}

The cyclotron resonance frequencies were fitted with the theoretical lineshape \cite{Kon95, Geo07, Kre07} (see Fig.~\ref{fig:TOF}). The measured frequencies were corrected for the count-rate-effect \cite{Kel03} whenever it was possible. For the lower statistics files, where the count-rate-class analysis was not possible, the statistical error was multiplied by a factor obtained from a comparison of the errors in the frequencies of all higher-statistics files with and without the count-rate-class analysis. The magnetic field $B$ at the time of measurement was interpolated from the well-known reference measurements before and after the measurement. The frequency ratio $r$ of the well-known reference ion to the ion of interest was determined (see Eq.~\ref{eq:r}). This ratio gives the mass ratio of the ion of interest to the reference ion (see Eq.~\ref{eq:rmref1}),

\begin{equation}
\label{eq:r}
\begin{split}
r &= \frac{\nu_{ref}}{\nu}~,\\
\end{split}
\end{equation}

\begin{equation}
\label{eq:rmref1}
\begin{split}
r &= \frac{m-m_e}{m_{ref}-m_e}~.\\
\end{split}
\end{equation}

In order to take into account fluctuations in the magnetic field, a correction of $\delta_B(\nu_{ref})/\nu_{ref}=5.7(8)\times 
10^{-11}\text{min}^{-1}\Delta t$, where $\Delta t$ is the time between the two reference measurements, was quadratically added to the statistical uncertainty of each frequency ratio. The weighted mean of the measured frequency ratios was calculated and used as the final value. The inner and outer errors \cite{Bir32} of the data sets were compared and the larger value of these two was taken as the error of the mean. Finally, the uncertainty due to mass-dependent shift $\delta_{m,lim}(r)/r=(7.5 \pm 0.4 \times 10^{-10}/u)\times \Delta m$ \cite{Elo09} and an additional residual relative uncertainty $\delta_{res,lim}(r)/r=7.9\times 10^{-9}$ \cite{Elo09} were quadratically added to the error.

\subsection{\label{sec:adjusted}Data evaluation}

In order to evaluate the masses of the measured nuclides, a least-squares adjustment was done in a similar manner as in 
Refs.~\cite{Aud86,AME,Muk08}. Here, we follow the notations used in those references. The input data $q_i$ for the least-squares method 
consist of the measured 17 frequency ratios between 13 nuclides (see Table~\ref{tab:ratios}) and the current mass values for the 13 nuclides from the Atomic Mass Evaluation 2003 (AME03) \cite{AME}. Thus, we have 30 input data to 13 nuclides involved in the frequency ratio measurements forming an overdetermined system. 

The input data of the thirteen AME03 values are simply $q_i=m_i\pm \delta m_i$. For the frequency ratios, a similar procedure as in 
Ref.~\cite{Muk08} was applied. Eq.~\ref{eq:rmref1} can be expressed as a linear equation in $m$:

\begin{equation}
\label{eq:rmref2}
\begin{split}
m - r\cdot m_{ref}&=m_e(1-r)~.\\
\end{split}
\end{equation}

In order to have the left side independent of the ratio $r$, a constant factor $C=A/A_{ref}$, where $A$ and $A_{ref}$ are the mass numbers 
of the reference ion and the ion of interest, is introduced. Then, a term $-C\cdot m_{ref}$ is added on both sides of Eq.~\ref{eq:rmref2}: 

\begin{equation}
\label{eq:rmref3}
\begin{split}
m - C\cdot m_{ref}&=(r-C)m_{ref}+m_e(1-r)~.\\
\end{split}
\end{equation}

Including the uncertainties $\delta r$, $\delta m_{ref}$, and $\delta m_e$, Eq.~\ref{eq:rmref3} yields:

\begin{equation}
\label{eq:linear_errors}
\begin{split}
m - C\cdot m_{ref}&=(r-C)m_{ref}+m_e(1-r)\\
&+ \{(r-C)\delta m_{ref}+m_{ref}\delta r\}~.\\
\end{split}
\end{equation}

In Eq.~\ref{eq:linear_errors}, the terms $\delta m_e(1-r)$ and $m_e\delta r$ have been neglected since they are small compared to 
$m_{ref}\delta r$. Since the left-side of Eq.~\ref{eq:linear_errors} is a continuous and differentiable function of $m$, a least-squares 
fit to this linear, overdetermined system  can be applied following Ref.~\cite{Aud86}. The measured data $q_i$ are obtained from 
Eq.~\ref{eq:qi} with the uncertainties $dq_i$ given in Eq.~\ref{eq:deltaqi}:

\begin{equation}
\label{eq:qi}
\begin{split}
q_i &=(r-C)m_{ref}+m_e(1-r)~,\\
\end{split}
\end{equation}

\begin{equation}
\label{eq:deltaqi}
\begin{split}
dq_i &= (r-C)\delta m_{ref}+m_{ref}\delta r~.\\
\end{split}
\end{equation}

Let the vector $\left|m\right\rangle$ represent the masses of 13 nuclides involved in the frequency ratio measurements and the vector $\left|q\right\rangle$ corresponds to the input data (17 rows obtained from Eq.~\ref{eq:qi} and 13 rows representing the AME03 mass values of the nuclides). Then, a $30\times 13$ matrix $K$ representing the coefficients $K\left|m\right\rangle=\left|q\right\rangle$ and a $30\times 30$ diagonal weight matrix $W$ with the elements $w_i^i=1/dq_i^2$ can be formed. The solution of the least-squares method yields a vector of adjusted masses $\left|\overline{m}\right\rangle=A^{-1}$$^tKW\left|q\right\rangle=R\left|q\right\rangle$ where $A^{-1}$ is the inverse of the normal matrix $A=~^tKWK$ which is a positive-definite and invertible square matrix of the order of $13$. The errors for the adjusted masses $\overline{m_i}$ are obtained as a square-root of the diagonal elements of the matrix $A^{-1}$. 

The adjusted data $\left|\overline{q}\right\rangle$ can be calculated as $\left|\overline{q}\right\rangle=KR\left|q\right\rangle$. Now, if the uncertainties $dq_i$ are very small for this overdetermined system (the number of input data $N_d=30$ $>$ the number of variables (masses) $N_v=13$), the normalized deviation between the adjusted data $\overline{q_i}$ and input data $q_i$ should have a Gaussian distribution with $\sigma=1$. For $N_d-N_v=30-13=17$ degrees of freedom, this gives a $\chi^2$ equal to 

\begin{equation}
\label{eq:chi2}
\begin{split}
\chi^2 &= \sum_{i=1}^{N_d}\left(\frac{\overline{q_i}-q_i}{dq_i}\right)^2~.\\
\end{split}
\end{equation}

The consistency can also be expressed as normalized $\chi$:

\begin{equation}
\label{eq:normchi}
\begin{split}
\chi_n &= \sqrt{\chi^2/(N_d-N_v)}
\end{split}
\end{equation}

for which the expected value is $1\pm 1/\sqrt{2(N_d-N_v)}$.

The influence of each datum $i$ on a mass $m_\nu$ can be seen from the $(i,\nu)$ element of a flow-of-information matrix $F=~^tR\otimes K$ 
($30\times 13$ matrix) \cite{Aud86}. Each column of $F$ represents all the contributions from all input data to a given mass $m_\nu$. The sum of these 
contributions is 1. The sum of influences along each row shows the significance of that datum.

\section{\label{sec:results}Results and discussion}

\subsection{\label{sec:ratios} Frequency ratios}
Altogether 20 frequency ratios were measured in this work (see Table~\ref{tab:ratios}). The number of measured frequency ratios is high because the reference nuclides in this mass region are known with quite a modest precision of about $0.6-2~\text{keV}$ or $\delta m/m \approx 1.1-4.1\times 10^{-8}$. Therefore, a small network of measurements provides more accurate mass values for the measured nuclides. In addition, some $Q_\text{EC}$ and $S_p$ values were measured directly to obtain a better precision.

\begin{figure}[!]
\centering
\rotatebox{270}{
\includegraphics[width=0.35\textwidth,clip]{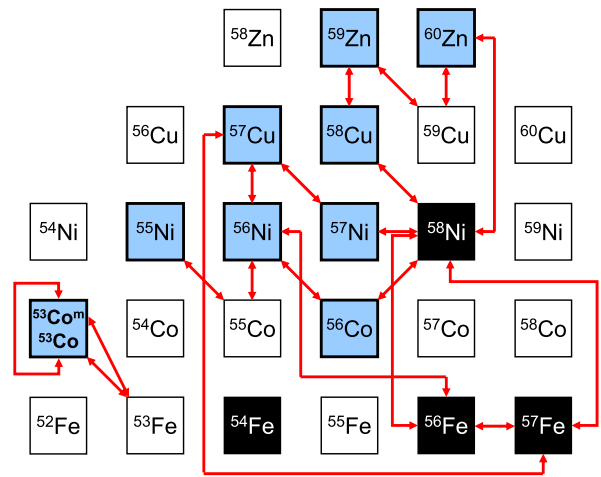}}
\caption{(Color online) The highlighted nuclides were measured in this work. The red arrows show the measured frequency ratio pairs.\label{fig:connections}}       
\end{figure}

\begin{table*}[!]
\caption{\label{tab:ratios} The measured frequency ratios ($r=\frac{\nu_{ref}}{\nu_c}$) for the nuclides. The references used, the 
production method, the number of measurements ($N_{meas}$) and the total number of ions in the resonances ($N_{ions}$) are also given in 
the table. Note that the frequency ratios of $^{56}$Co and $^{57}$Ni relative to $^{58}$Ni were measured in the HIGISOL run and the last three frequency ratios in an off-line run employing an electric discharge ion source. Uncertainties in the frequency ratios are given without ($\delta r$) and with an additional relative residual uncertainty of $7.9\times 10^{-9}$ \cite{Elo09} ($\delta r_{all}$).}
\begin{ruledtabular}
\begin{tabular}{llllll}
Nuclide & Ref. & Prod. Method & $N_{meas}$ & $N_{ions}$ & $r(\delta r)(\delta r_{all})$\\
\hline
$^{53}$Co & $^{53}$Fe & 40 MeV p on $^{54}$Fe & 9 & $30731$ & $1.000 \ 168 \ 055 \ 7(46)(92)$\\
$^{53}$Co$^m$ & $^{53}$Fe & 40 MeV p on $^{54}$Fe & 6 & $5152$ & $1.000 \ 232 \ 415(23)(24)$\\
$^{53}$Co$^m$ & $^{53}$Co & 40 MeV p on $^{54}$Fe & 4 & $3122$ & $1.000 \ 064 \ 357(27)(28)$\\
$^{55}$Ni & $^{55}$Co & 25 MeV $^3$He$^{2+}$ on $^{54}$Fe & 6 & $5690$ & $1.000 \ 169 \ 879 \ 6(81)(113)$\\
$^{56}$Ni & $^{55}$Co & 40 MeV p on $^{58}$Ni & 4 & $12644$ & $1.018 \ 203 \ 600 \ 5(36)(88)$\\
$^{56}$Ni & $^{56}$Co & 25 MeV $^3$He$^{2+}$ on $^{54}$Fe & 4 & $9861$ & $1.000 \ 040 \ 930 \ 2(39)(88)$\\
$^{56}$Ni & $^{56}$Fe & 50 MeV p on $^{58}$Ni & 4 & $24970$ & $1.000 \ 128 \ 579 \ 3(48)(92)$\\
$^{57}$Cu & $^{56}$Ni & 40 MeV p on $^{58}$Ni & 6 & $9460$ & $1.018 \ 002 \ 434 \ 0(57)(99)$\\
$^{57}$Cu & $^{57}$Ni & 40/50 MeV p on $^{58}$Ni & 11 & $10895$ & $1.000 \ 165 \ 446 \ 9(56)(97)$\\
$^{57}$Cu & $^{57}$Fe & 40 MeV p on $^{58}$Ni & 5 & $8020$ & $1.000 \ 242 \ 695(13)(15)$\\
$^{58}$Cu & $^{58}$Ni & 40 MeV p on $^{58}$Ni & 7 & $20384$ & $1.000 \ 158 \ 637 \ 1(33)(86)$\\
$^{59}$Zn & $^{58}$Cu & 25 MeV $^3$He$^{2+}$ on $^{58}$Ni & 2 & $1399$ & $1.017 \ 340 \ 531(21)(22)$\\
$^{59}$Zn & $^{59}$Cu & 25 MeV $^3$He$^{2+}$ on $^{58}$Ni& 5 & $6741$ & $1.000 \ 166 \ 532 \ 0(93)(122)$\\
$^{60}$Zn & $^{58}$Ni & 25 MeV $^3$He$^{2+}$ on $^{58}$Ni & 6 & $15744$ & $1.034 \ 633 \ 758 \ 6(49)(95)$\\
$^{60}$Zn & $^{59}$Cu & 25 MeV $^3$He$^{2+}$ on $^{58}$Ni & 5 & $9786$ & $1.017 \ 006 \ 490 \ 0(48)(94)$\\
$^{56}$Co & $^{58}$Ni & 105 MeV $^{20}$Ne$^{4+}$ on $^{nat}$Ca & 6 & $21159$ & $0.965 \ 556 \ 038 \ 7(54)(94)$\\
$^{57}$Ni & $^{58}$Ni & 75 MeV $^{20}$Ne$^{4+}$ on $^{nat}$Ca  & 5 & $5726$ & $0.982 \ 816 \ 024 \ 9(96)(124)$\\
$^{56}$Fe & $^{58}$Ni & discharge ion source & 20 & $119485$ & $0.965 \ 471 \ 417 \ 0(26)(81)$\\
$^{57}$Fe & $^{56}$Fe & discharge ion source& 14 & $55197$ & $1.017 \ 886 \ 256 \ 4(21)(83)$\\
$^{57}$Fe & $^{58}$Ni & discharge ion source & 21 & $98650$ & $0.982 \ 740 \ 085 \ 8(17)(80)$\\
\end{tabular}
\end{ruledtabular}
\end{table*}

\subsection{\label{sec:mass} Mass excess values}
The nuclides other than $^{53}$Co and $^{53}$Co$^m$, formed a network of $13$ nuclides and $17$ measured frequency ratios. For these nuclides, a least-squares method described in Sec.~\ref{sec:adjusted} was applied and adjusted mass values were obtained. The normalized $\chi=1.08$ was well within the expected value $1.00 \pm 0.17$, and therefore, no additional error was added to the frequency ratios. The biggest contribution to the $\chi^2$ value ($27~\%$) comes from the $^{58}$Cu AME03 mass value which is $3.6(17)~\text{keV}$ higher than the adjusted value obtained with the JYFLTRAP results. In addition to $^{58}$Cu, also the AME03 values of $^{55}$Co (13~\%), $^{60}$Zn (10~\%), $^{59}$Zn (8~\%)and $^{58}$Ni (7~\%) have a substantial contribution to the $\chi^2$ value. This is also seen in the adjusted values which deviate from the AME03 values of these nuclides. 

In the following, the mass excess results for the radioactive nuclides are compared to earlier experiments (see Figs.~\ref{fig:co_me}, \ref{fig:ni_me}, \ref{fig:cu_me}, and \ref{fig:zn_me}) and discussed nuclide by nuclide. The results for the nuclides mainly used as references are also summarized (see Figs.~\ref{fig:co_refs}, \ref{fig:fe_refs}, and \ref{fig:ni_refs}). The directly measured values were used for $^{53}$Co and $^{53}$Co$^m$. For the rest, the adjusted mass values (see Table~\ref{tab:MEadj}) were applied. The results of $^{53}$Co and $^{53}$Co$^m$ include also a new value for the excitation energy of the high-spin isomer.

\begin{table*}[!]
\caption{\label{tab:MEadj} Mass excess values ($ME$) and a comparison to literature values ($ME_{AME}$) \cite{AME}. The mass excess values are the adjusted values except for $^{53}$Co and $^{53}$Co$^m$ which were not included in the network.} 
\begin{ruledtabular}
\begin{tabular}{llllll}
Nuclide & $ME$ (keV) & $ME_{AME}$ (keV) & $ME-ME_{AME}$ (keV) & Input & Influence (\%)\\
\hline
$^{56}$Fe & -60605.38(37) & -60605.4(7) & -0.03(78) & $^{56}\text{Fe}-^{58}$Ni & 28.4\\ 
		 & & & 	&  $^{57}\text{Fe}-^{56}$Fe & 24.9\\ 
		 & & & 	& $^{56}\text{Ni}-^{56}$Fe & 18.1\\
		 & & & 	&  $^{56}$Fe, AME03 & 28.6\\ 

$^{57}$Fe & -60179.78(38) &  -60180.1(7) &	0.35(78) & $^{57}\text{Fe}-^{56}$Fe & 28.1\\
	 	 & & & 	&  $^{57}\text{Fe}-^{58}$Ni & 32.8\\ 
		 & & & 	&  $^{57}\text{Cu}-^{57}$Fe & 8.4\\
		 & & & 	&  $^{57}$Fe, AME03 & 30.8\\ 	 

$^{53}$Co\footnotemark[1] & -42657.3(15) & -42645(18) & -13(18) & $^{53}\text{Co} - ^{53}$Fe & 65.6 \\
 &  &  & & $^{53}\text{Co} - ^{53}$Co$^m - ^{53}$Fe & 34.4 \\
$^{53}$Co$^m$\footnotemark[1] & -39482.9(16) & -39447(22) & -36(22) & $^{53}\text{Co}^m - ^{53}$Fe & 53.9 \\
 &  &  & & $^{53}$Co$^m - ^{53}$Co$ - ^{53}$Fe & 46.1 \\
 
$^{55}$Co & -54028.72(48) &	-54027.6(7) &	-1.16(87) & $^{55}$Ni$-^{55}$Co & 0.2\\
 & & &	 & $^{56}$Ni$-^{55}$Co  & 56.1 \\ 		
 & & &	 & $^{55}$Co, AME03  & 43.7  \\
 
$^{56}$Co & -56038.81(47) & -56039.4(21) &	0.5(22) & $^{56}$Co$-^{58}$Ni & 50.9 \\ 
 & & &	 & $^{56}$Ni$-^{56}$Co   &  44.2\\	
 & & &	  & $^{56}$Co, AME03 & 4.9 \\  
 
$^{55}$Ni & -45334.69(75) &  	-45336(11)	& 0.9(110) & $^{55}$Ni$-^{55}$Co & 99.5 \\	 
 & & &	  & $^{55}$Ni, AME03 & 0.5\\  
 
$^{56}$Ni & -53906.02(42) & 	-53904(11) &	-2.3(111) & $^{56}$Ni$-^{55}$Co  & 23.3 \\		
 & & &	 &  $^{56}$Ni$-^{56}$Co & 22.6 \\
 & & &	 &  $^{56}$Ni$-^{56}$Fe & 35.8 \\
 & & &	 &  $^{57}$Cu$-^{56}$Ni & 18.2 \\
 & & &	 &  $^{56}$Ni, AME03 & 0.1 \\ 
  
$^{57}$Ni & -56082.11(55) &	-56082.0(18) &	-0.1(19) & $^{57}$Ni$-^{58}$Ni & 46.5\\ 
 & & &	&   $^{57}$Cu$-^{57}$Ni & 44.4 \\
 & & &	&   $^{57}$Ni, AME03 & 9.1 \\		

$^{58}$Ni & -60226.96(35) & -60227.7(6) &	0.74(70)	& $^{56}$Fe$-^{58}$Ni & 17.8\\ 	  
	 & & & &  $^{57}$Fe$-^{58}$Ni & 18.3\\
	 & & & & $^{56}$Co$-^{58}$Ni & 8.8\\	
	 & & & &  $^{57}$Ni$-^{58}$Ni  & 5.5 \\
	 & & & &  $^{58}$Cu$-^{58}$Ni  & 6.5 \\	
	 & & & &  $^{60}$Zn$-^{58}$Ni & 10.2 \\	
 & & &	&  $^{58}$Ni, AME03 & 33.0 \\	 
 
$^{57}$Cu & -47307.20(50) &	-47310(16) &	2(16) & $^{57}$Cu$-^{56}$Ni & 46.0\\		
 & & &	&  $^{57}$Cu$-^{57}$Ni & 27.3\\ 
  & & &	&  $^{57}$Cu$-^{57}$Fe & 26.6\\
   & & &	&  $^{57}$Cu, AME03 & 0.1\\	 
	 
$^{58}$Cu & -51665.69(52) & -51662.1(16) & -3.6(17) & $^{58}$Cu$-^{58}$Ni & 78.2\\	
 & & &	&   $^{59}$Zn$-^{58}$Cu & 10.5 \\		
  & & &	&   $^{58}$Cu, AME03 & 11.3 \\			
 	 
$^{59}$Cu & -56356.83(54) & -56357.2(8) &	0.40(95) & $^{59}$Zn$-^{59}$Cu & 10.8\\
& & &	&  $^{60}$Zn$-^{59}$Cu  & 42.4\\
& & &	&  $^{59}$Cu, AME03  & 46.8\\		 		

$^{59}$Zn  & -47213.93(74) &	-47260(40) & 47(40) & $^{59}$Zn$-^{58}$Cu  & 29.8\\
	 & & &	 & $^{59}$Zn$-^{59}$Cu  & 70.2\\	
& & &	&  $^{59}$Zn, AME03  & 0.04\\		 

$^{60}$Zn & -54172.67(53) &	-54188(11) & 15(11) & $^{60}$Zn$-^{58}$Ni & 65.9\\
& & &	&  $^{60}$Zn$-^{59}$Cu  & 33.8 \\	
& & &	&  $^{60}$Zn, AME03  & 0.2\\																						
\end{tabular}
\end{ruledtabular}
\footnotetext[1]{The mass excess value has been determined with respect to the reference nucleus $^{53}$Fe either directly or via ground or isomeric state for the isomeric or ground state, respectively. A weighted mean of these two values has been adopted for the mass excess value.}
\end{table*}

\subsubsection{\label{sec:com53} \textbf{$^{53}$Co and the spin-gap isomer in $^{53}$Co}}

\begin{figure}[!]
\centering
\includegraphics[width=0.45\textwidth,clip]{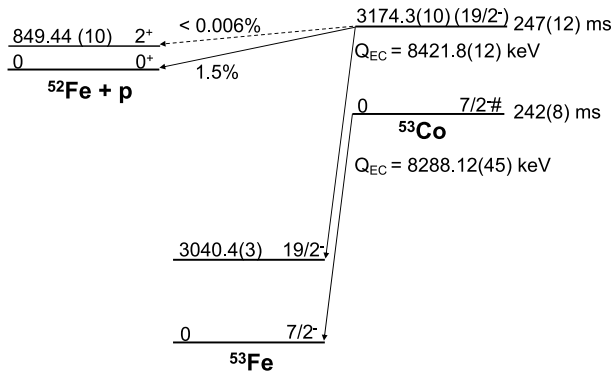}
\caption{Revised decay scheme of $^{53}$Co. For the $Q_\text{EC}$ values, see Sec.~\ref{sec:qec}.}
\label{fig:53cofig}       
\end{figure}

\begin{figure}[!]
\center{
\rotatebox{270}{
\resizebox{0.45\textwidth}{!}{
\includegraphics{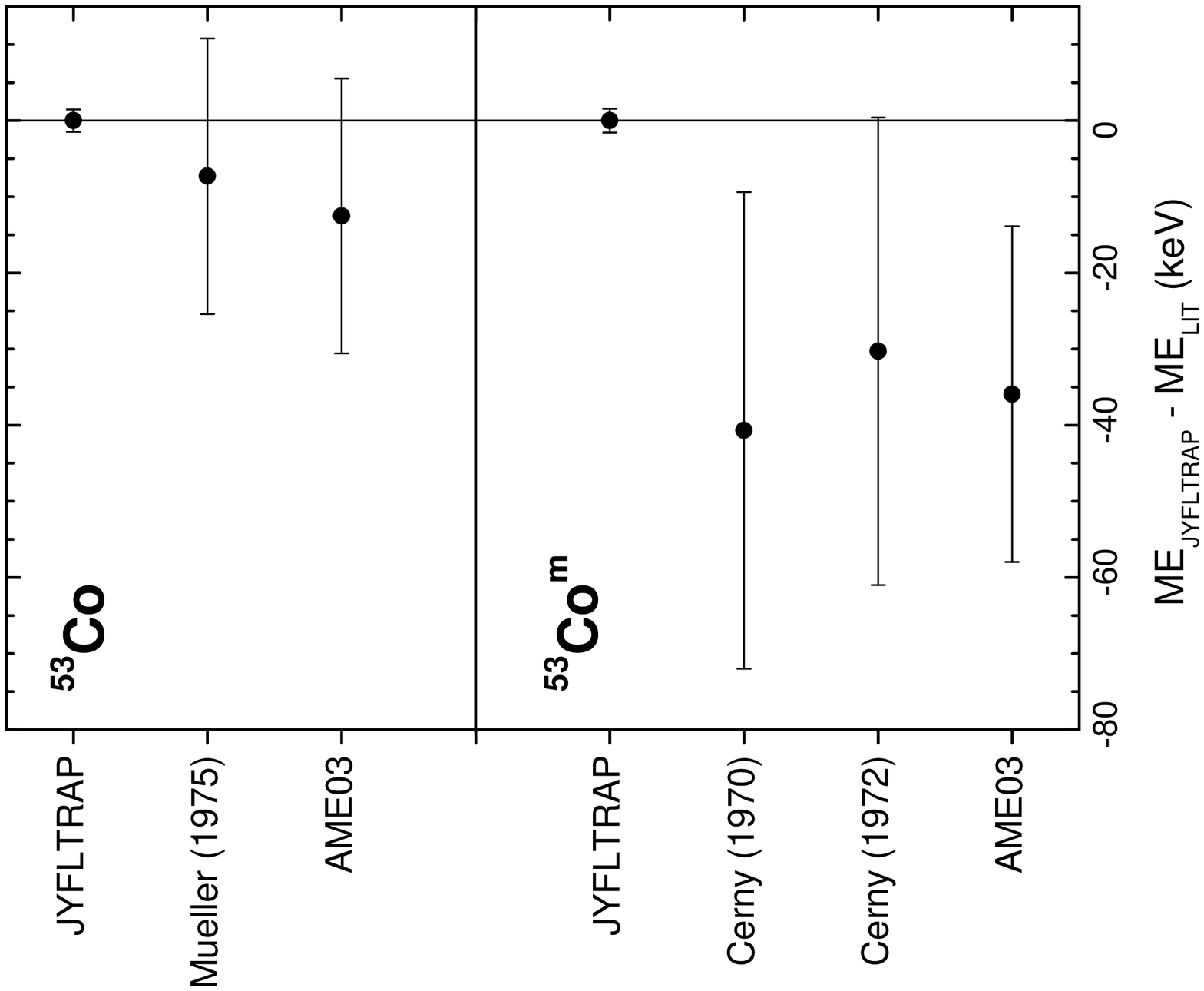}}}
\caption{Differences between the experimental mass excess values of the ground and isomeric states of $^{53}$Co measured at JYFLTRAP with 
respect to the earlier experiments \cite{Mue75,Cer70,Cer72} and AME03 \cite{AME}.}
\label{fig:co_me}       
}
\end{figure}

The ground state mass of $^{53}$Co in AME03 is based on the measured $Q$ value of the $^{58}$Ni(p,$^6$He)$^{53}$Co reaction \cite{Mue75} which is in agreement with the new JYFLTRAP value. Proton decay of the spin-gap isomer $^{53}\text{Co}^m$ was observed in Refs.~\cite{Cer70,Cer72}. The observed proton peak energies $E_{lab}=1570(30)~\text{keV}$ \cite{Cer70} and $E_{CM}=1590(30)~\text{keV}$ \cite{Cer72} and the tabulated mass of $^{52}$Fe \cite{AME} result in an excitation energy of $3197(29)~\text{keV}$ and a mass excess value of $-39447(22)~\text{keV}$ for $^{53}$Co$^m$. The new JYFLTRAP mass excess value for the isomer agrees with the one from Ref.~\cite{Cer72} but disagrees with the value of Ref.~\cite{Cer70} and the adopted AME03 value \cite{AME} (see Fig.~\ref{fig:co_me}). 

The excitation energy of the isomer was measured directly in the $^{53}\text{Co}^{m}-^{53}$Co pair yielding an energy of  $3174.5(14)~\text{keV}$. It was also determined indirectly from the energy difference of the $^{53}\text{Co}^{m}-^{53}$Fe and $^{53}\text{Co}-^{53}$Fe pairs resulting in an excitation energy of $3174.1(13)~\text{keV}$. The weighted average of these results gives an excitation energy of $3174.3(10)~\text{keV}$. With the $^{52}$Fe mass excess from Ref.~\cite{AME}, this would correspond to a proton peak energy of $E_{lab}=1530(7)~\text{keV}$. A new decay scheme for $^{53}$Co based on this work is presented in Fig.~\ref{fig:53cofig}. 

Coulomb energy differences ($CED$) show the differences in the excitation energies between excited isobaric analog states (IAS) with increasing spin. The isobaric analog state of the $19/2^-$ isomer at $E_x=3174.3(10)~\text{keV}$ in $^{53}$Co lies at $3040.4(3)~\text{keV}$ in $^{53}$Fe. This yields a CED of $133.9(10)~\text{keV}$ which  improves the precision considerably compared to the AME03 value of $157(29)~\text{keV}$. The new excitation energy for the isomer is quite close to the erroneous excitation energy of $3179(30)~\text{keV}$ adopted accidentally in Ref.~\cite{Wil03} instead of the AME03 value of $3197(29)~\text{keV}$ \cite{AME}. Thus, the new result for the CED is (by chance) in agreement with the result of Ref.~\cite{Wil03} where a smooth rise of CED was observed from the ${7/2^-}$ state to the ${19/2^-}$ isomeric state. This smooth rise reflects the gradual alignment of the $\nu(f_{7/2})^{-2}$ pair from $J=0$ to $J=6$ in $^{53}$Co (for the $\pi(f_{7/2})^{-2}$ pair in $^{53}$Fe) \cite{Wil03,Ben07}.

\subsubsection{\label{sec:55ni}\textbf{$^{55}$Ni}}

The mass of $^{55}$Ni has been previously measured via $^{58}\text{Ni}(^3\text{He},^6\text{He})^{55}\text{Ni}$ reactions at the Michigan State 
University in the 1970s \cite{Pro72,Mue75,Mue77} and via a $\beta$-endpoint measurement conducted at IGISOL \cite{Ays84}. The AME03 mass 
excess value is based on the $Q$ value of Ref.~\cite{Mue75} corrected by a new $Q$ value for the calibration reaction 
$^{27}$Al($^3$He,$^6$He)$^{24}$Al used in Ref.~\cite{Mue77}. The new JYFLTRAP value agrees with all the other values except with Ref.~\cite{Pro72} for which the $Q$ values used in the energy calibration are not given (see Fig.~\ref{fig:ni_me}).

\subsubsection{\label{sec:56ni}\textbf{$^{56}$Ni}}
\begin{figure}[!]
\center{
\rotatebox{270}{
\resizebox{0.45\textwidth}{!}{
\includegraphics{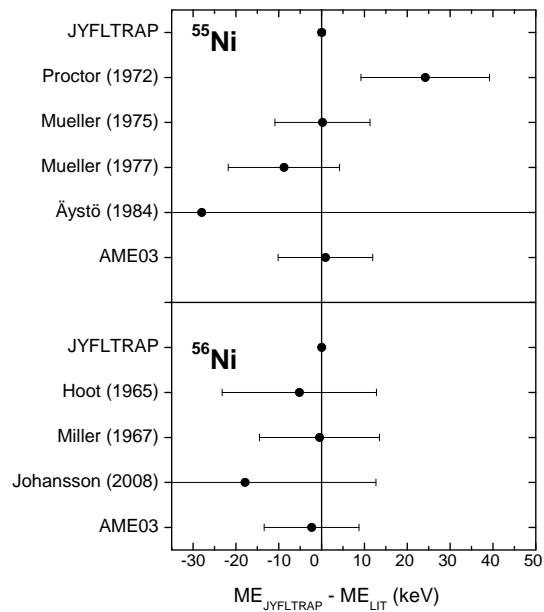}}}
\caption{Differences between the experimental mass excess values of the nickel isotopes measured at JYFLTRAP with respect to the earlier 
experiments \cite{Pro72,Mue75,Mue77,Ays84,Hoo65,Mil67,Joh08} and AME03 \cite{AME}.}
\label{fig:ni_me}       
}
\end{figure}

The current mass excess value of $^{56}$Ni is based on the $Q$ values of the reactions $^{58}$Ni(p,t)$^{56}$Ni \cite{Hoo65} and 
$^{54}$Fe($^3$He,n)$^{56}$Ni \cite{Mil67}. Recently, prompt proton decay ($E_p=2540(30)~\text{keV}$) was observed from a level at 
$9735(2)~\text{keV}$ in a rotational band of $^{56}$Ni  \cite{Joh08}. The JYFLTRAP value agrees with all previous experiments (see Fig.~\ref{fig:ni_me}) but is 26 times more accurate than the adopted value.

\subsubsection{\label{sec:57cu}\textbf{$^{57}$Cu}}

The mass of $^{57}$Cu has been earlier determined via $\beta$-endpoint energy \cite{Shi84}, the $Q$ values of 
$^{58}$Ni($^7$Li,$^8$He)$^{57}$Cu measured at the National Superconducting Cyclotron Laboratory \cite{She85} and at the Texas A \& M 
cyclotron \cite{Gag86} and the $Q$ value of $^{58}$Ni($^{14}$N,$^{15}$C)$^{57}$Cu \cite{Sti87}. The JYFLTRAP value is 31 times more accurate than the adopted AME03 value and in agreement with these measurements (see Fig.~\ref{fig:cu_me}).

\subsubsection{\label{sec:58cu}\textbf{$^{58}$Cu}}

The mass of $^{58}$Cu was earlier based on the measurements of the threshold energy for the reaction $^{58}$Ni$(p,n)^{58}$Cu \cite{Fre65,Bon66,Ove69}. The $Q_\text{EC}$ value for $^{58}$Cu has been measured at JYFLTRAP, $Q_{EC}=8555(9)~\text{keV}$ \cite{Kol04} which yields a mass excess of $-51673(9)~\text{keV}$ when using the AME03 value for $^{58}$Ni. The new mass excess value of $-51665.69(52)~\text{keV}$ disagrees with the $(p,n)$ threshold energies and with the AME03 value but is in agreement with the previous JYFLTRAP result \cite{Kol04} and the result derived from prompt proton emission from $^{58}$Cu \cite{Rud02}. 

\begin{figure}[!]
\center{
\rotatebox{270}{
\resizebox{0.45\textwidth}{!}{
\includegraphics{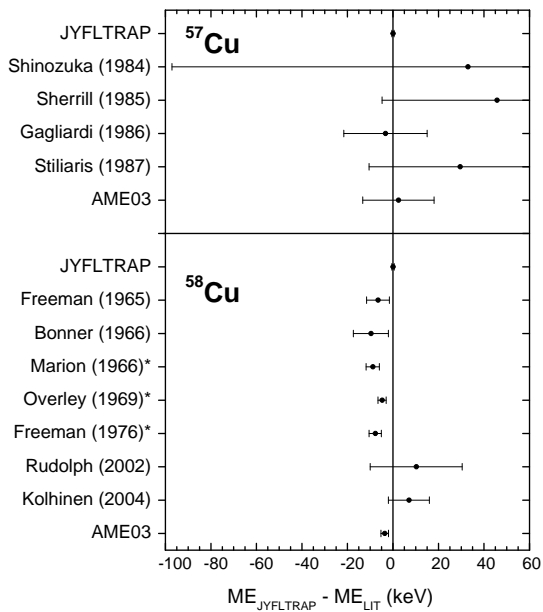}}}
\caption{Differences between the experimental mass excess values of the copper isotopes measured at JYFLTRAP with respect to the earlier 
experiments \cite{Shi84,She85,Gag86,Sti87,Fre65,Bon66,Mar66,Ove69,Fre76,Rud02,Kol04} and AME03 \cite{AME}. The values marked with $*$ are only recalculated values from previous $(p,n)$ measurements of \cite{Fre65} and \cite{Bon66}.}
\label{fig:cu_me}       
}
\end{figure}

The problems in the determination of the $Q$ values from the threshold energies explain the discrepancy between the results. Freeman \cite{Fre76} has suggested that if threshold energies are used to derive $Q$ values for mass determination, the errors should be increased by some, albeit arbitrary, factor ($\sqrt{2}$ or $2$). In addition, Refs.~\cite{Ove69} and \cite{Mar66} only recalculate the values measured in Refs.~\cite{Fre65,Bon66}. Thus, Ref.~\cite{Ove69} should not be averaged with Ref.~\cite{Bon66}. A revised value is given in Ref.~\cite{Fre76}. However, none of these values agree with JYFLTRAP (see Fig.~\ref{fig:cu_me}). A similar deviation of $-5.3(39)~\text{keV}$ is observed when the JYFLTRAP value for the $^{54}$Co mass \cite{Ero08} is compared with the threshold energy for the reaction $^{54}$Fe$(p,n)^{54}$Co \cite{Fre76}. 

\subsubsection{\label{sec:59zn}\textbf{$^{59}$Zn}}

The JYFLTRAP mass excess value for $^{59}$Zn agrees with the mass derived from the $Q_\text{EC}$ value of Ref.~\cite{Ara81} and almost agrees 
with the value derived from the $^{58}$Ni(p,$\pi^-$)$^{59}$Zn $Q$ value \cite{She83}. However, the AME03 value deviates 
from the JYFLTRAP value slightly more than $1\sigma$ (see Fig.~\ref{fig:zn_me}).

\subsubsection{\label{sec:60zn}\textbf{$^{60}$Zn}}

\begin{figure}[!]
\center{
\rotatebox{270}{
\resizebox{0.45\textwidth}{!}{
\includegraphics{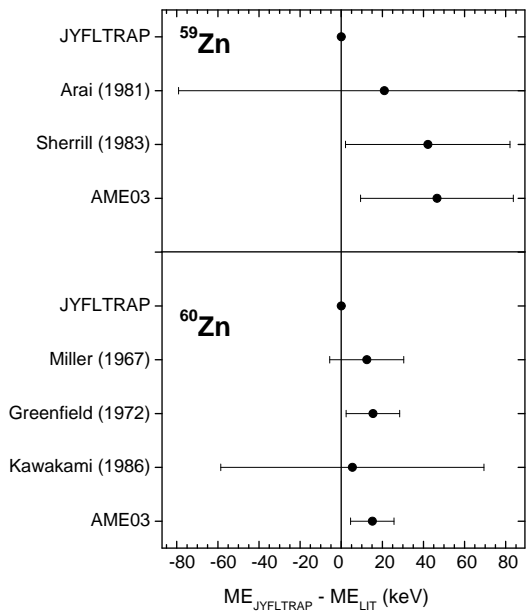}}}
\caption{Differences between the experimental mass excess values of the zinc isotopes measured at JYFLTRAP with respect to the earlier 
experiments \cite{Ara81,She83,Mil67,Gre72,Kaw86} and AME03 \cite{AME}.}
\label{fig:zn_me}       
}
\end{figure}

The mass of $^{60}$Zn is based on the $Q$ values for the reaction $^{58}$Ni($^3$He,n)$^{60}$Zn \cite{Mil67,Gre72} in the AME03 
compilation. The new JYFLTRAP value agrees with the one from Ref.~\cite{Mil67} but disagrees slightly with Ref.~\cite{Gre72} and with the AME03 value. 
The $Q_\text{EC}$ value for the beta decay of $^{60}$Zn \cite{Kaw86} is in agreement with the mass excess value measured in this work~(see 
Fig.~\ref{fig:zn_me}).

\subsubsection{\label{sec:refs}\textbf{References $^{56,57}$Fe, $^{55,56}$Co, $^{57,58}$Ni, and $^{59}$Cu}}

\begin{figure}[!]
\center{
\rotatebox{270}{
\resizebox{0.45\textwidth}{!}{
\includegraphics{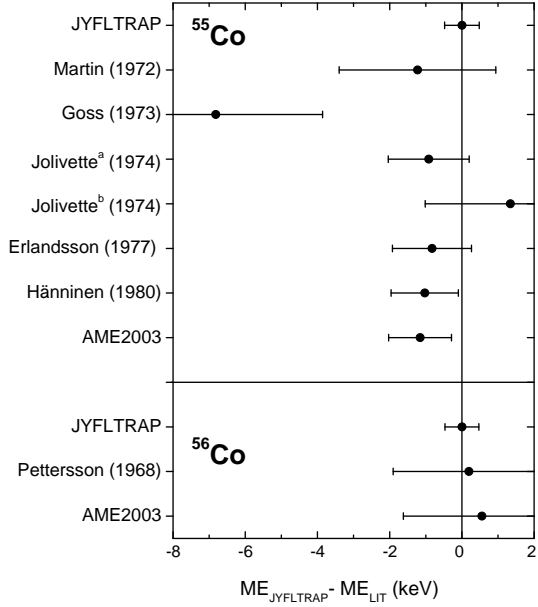}}}
\caption{Differences between the adjusted mass excess values of the cobalt isotopes used as references with respect to the earlier experiments \cite{Mar72,Gos73,Jol74,Erl77,Han80,Pet68} and AME03 \cite{AME}. $^{55}$Co is discussed in the text and the $^{56}$Co value of \cite{Pet68} is based on a beta-endpoint energy. Jolivette$^a$ refers to the $^{58}$Ni(p,$\alpha$)$^{55}$Co $Q$ value \cite{Jol74} and Jolivette$^b$ to the $^{54}$Fe($^3$He,d)$^{55}$Co $Q$ value \cite{Jol74}.}
\label{fig:co_refs}       
}
\end{figure}

Of the nuclides used as references in the frequency ratio measurements, $^{56}$Ni and $^{58}$Cu, have already been discussed above. Here, we concentrate on other reference nuclides: $^{56,57}$Fe, $^{55,56}$Co, $^{57,58}$Ni, and $^{59}$Cu. Masses of these reference nuclides close to $^{56}$Ni are known with a rather modest precision of $0.6-2.1~\text{keV}$. However, with the network of mass measurements, the precisions of these mass excess values have been improved to $0.35-0.55~\text{keV}$. The adjusted mass excess values for the used references agree well with the earlier results except for $^{55}$Co which deviates $-1.16(87)~\text{keV}$ from the AME03 value and for $^{58}$Ni for which the deviation is $0.74(70)~\text{keV}$. The deviation at $^{55}$Co is also seen in the mass excess values of $^{56}$Ni with respect to different reference nuclides. The mass excess value obtained for $^{56}$Ni with the $^{55}$Co reference is significantly higher than the values obtained with $^{56}$Co and $^{56}$Fe references suggesting that $^{55}$Co might have a too high mass excess value in Ref.~\cite{AME}. The former values of $^{55}$Co are based on $^{54}$Fe(p,$\gamma$)$^{55}$Co \cite{Mar72, Erl77, Han80}, $^{58}$Ni(p,$\alpha$)$^{55}$Co \cite{Gos73, Jol74} and $^{54}$Fe($^3$He,d)$^{55}$Co \cite{Jol74}. Of these, only the first $^{54}$Fe(p,$\gamma$)$^{55}$Co value \cite{Mar72} and the $^{54}$Fe($^3$He,d)$^{55}$Co \cite{Jol74} agree with JYFLTRAP (see Fig.~\ref{fig:co_refs}). The rest seem to overestimate the mass excess value.

\begin{figure}[!]
\center{
\rotatebox{270}{
\resizebox{0.45\textwidth}{!}{
\includegraphics{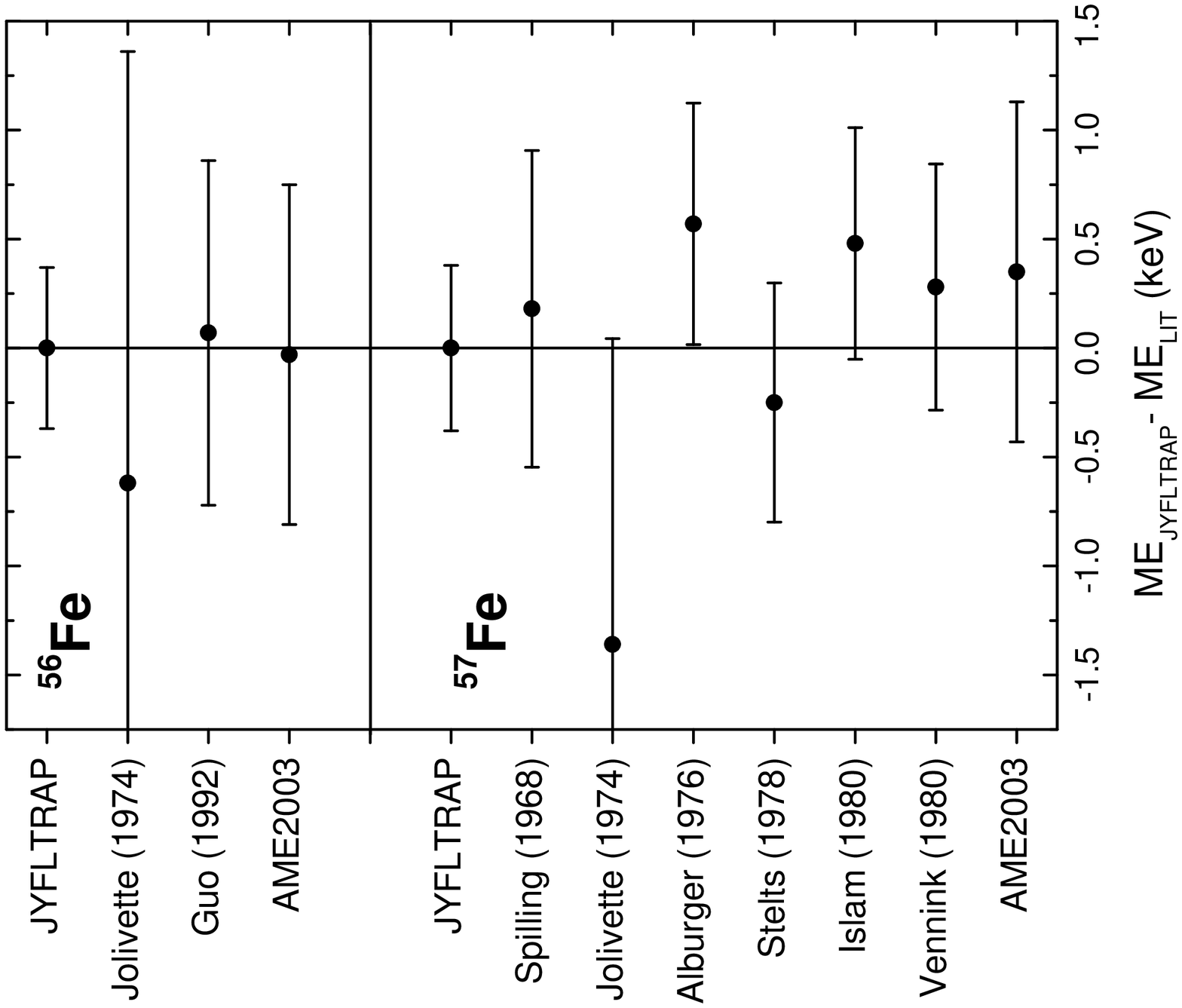}}}
\caption{Differences between the adjusted mass excess values of the iron isotopes used as references with respect to the earlier experiments and AME03 \cite{AME}. The earlier measurements are based on $^{59}$Co(p,$\alpha$)$^{56}$Fe \cite{Jol74} and $^{55}$Mn(p,$\gamma$)$^{56}$Fe \cite{Guo92} for $^{56}$Fe and on $^{56}$Fe(n,$\gamma$)$^{57}$Fe \cite{Spi68, Alb76, Ste78, Isl80, Ven80} and $^{56}$Fe(d,p)$^{57}$Fe \cite{Jol74}.}
\label{fig:fe_refs}       
}
\end{figure}

\begin{figure}[!]
\center{
\rotatebox{270}{
\resizebox{0.45\textwidth}{!}{
\includegraphics{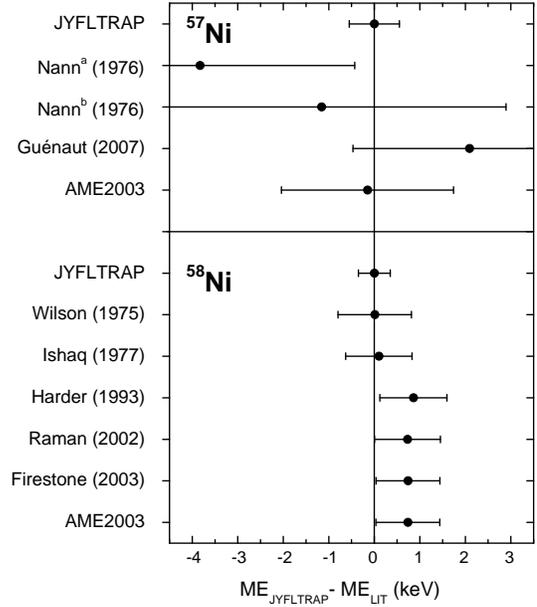}}}
\caption{Differences between the adjusted mass excess values of the nickel isotopes used as references with respect to the earlier experiments and AME03 \cite{AME}. The previous measurements are based on the $Q$ values of $^{59}$Ni(p,t)$^{57}$Ni (Nann$^a$) \cite{Nan76} and $^{58}$Ni($^3$He,$\alpha$)$^{57}$Ni (Nann$^b$) \cite{Nan76} and the frequency ratio for $^{57}$Ni-$^{85}$Rb \cite{Gue07} for $^{57}$Ni and the $Q$ values for $^{58}$Ni(n,$\gamma$)$^{59}$Ni  \cite{Wil75, Ish77, Har93, Ram02, Fir03} for $^{58}$Ni.}
\label{fig:ni_refs}       
}
\end{figure}

For the reference $^{58}$Ni, the earlier (n,$\gamma$) measurements \cite{Wil75, Ish77} agree almost perfectly with the JYFLTRAP value whereas the newer mass excess values together with the AME03 value disagree with it by $1\sigma$. Otherwise the iron and nickel reference nuclides agree surprisingly well with the earlier experiments (see Figs.~\ref{fig:fe_refs} and \ref{fig:ni_refs}) although many of these results have been measured precisely via (n,$\gamma$) reactions. This comparison shows that the uncertainties in the JYFLTRAP values are at a reasonable level. In addition, we could determine the neutron separation energies for $^{57}$Fe and $^{58}$Ni directly resulting in $S_n=7645.8(4)~\text{keV}$ and  $S_n=12216.4(7)~\text{keV}$ in agreement with the AME03 values $S_n=7646.10(3)~\text{keV}$ and $S_n=12217.0(18)~\text{keV}$ for $^{57}$Fe and $^{58}$Ni, respectively. The adjustment procedure improved also the precision of the radioactive $^{59}$Cu from $0.8~\text{keV}$ to $0.54~\text{keV}$. The new value agrees well with the AME03 value based on $^{58}$Ni(p,$\gamma$)$^{59}$Cu reactions \cite{Bon63,Fod70,Kla75}.

\subsection{\label{sec:qec}$Q_\text{EC}$ values and mirror decays}

The $Q_\text{EC}$ values are directly obtained by measuring the frequency ratio $r$ between the beta-decay mother (mass $m_m$) and daughter 
(mass $m_d$) in a Penning trap:

\begin{equation}
\label{eq:qec}
\begin{split}
Q_{EC} &= (m_m-m_d)c^2\\
	&= (r-1)(m_d-m_e)c^2~.\\
\end{split}
\end{equation}

With this method, the $Q_\text{EC}$ values can be determined to high precision even if the reference (daughter) nuclide has a moderate 
precision. The mass excesses for the daughter nuclides were taken from the adjusted mass values (Table~\ref{tab:MEadj}).

The $Q_\text{EC}$ values are tabulated in Table~\ref{tab:qec}. The mirror-decay $Q_\text{EC}$ values of $T=1/2$ nuclides $^{53}$Co, $^{55}$Ni, $^{57}$Cu, and $^{59}$Zn as well as the $Q_\text{EC}$ values for the $T_Z=0$ nuclides $^{56}$Ni and $^{58}$Cu in the $T=1$ triplets at $A=56$ and $A=58$ were directly determined from the frequency ratio measurements against their beta-decay daughters. In addition, the $Q_\text{EC}$ value for the spin-gap isomer $^{53}$Co$^m$ was measured relative to the $^{53}$Fe ground state. $^{53}$Co$^m$ decays dominantly to its isobaric analogue state (IAS) at $3040.4(3)~\text{keV}$ in $^{53}$Fe \cite{NDS87} for which a $Q_\text{EC}$ value of $8421.8(12)~\text{keV}$ is obtained. The $Q_\text{EC}$ value for $^{60}$Zn was determined from the adjusted mass value for $^{60}$Zn and the AME03 value for $^{60}$Cu \cite{AME}.

Recently, corrected $ft$ values, $\mathcal{F}t$, have been calculated for $T=1/2$ mirror transitions up to $^{45}$V \cite{Sev08,Nav08}. The $Q_\text{EC}$ values measured in this work offer a possibility to expand these studies from $^{53}$Co up to $^{59}$Zn. Table~\ref{tab:ft} summarizes the current averages of half-lives and branching ratios as well as electron-capture probabilities needed to calculate the $ft$ value. Experimental Gamow-Teller matrix elements  $\left|\left\langle \sigma\tau\right\rangle\right|$ have been calculated from the Gamow-Teller strength $B(GT)$:

\begin{equation}
\label{eq:GT}
\begin{split}
B(GT) &= \frac{C}{ft}-B(F) \\
\left\langle \sigma\tau\right\rangle^2&=\frac{B(GT)}{(g_A/g_V)^2}\\
\end{split}
\end{equation}
 
where the constant $C=2\cdot \overline{\mathcal{F}t}^{0^+\rightarrow0^+}=6143.5(17)~\text{s}$ \cite{Tow10}, $B(F)$ is the Fermi strength, which equals $1$ for $T=1/2$ mirror decays, and $g_A/g_V=-1.2695(29)$ \cite{PDG} is the ratio of the axial vector to the vector coupling constant. Isospin symmetry breaking and radiative corrections have not been taken into account. Their effect would be less than $1~\%$ of the $ft$ value which is small compared to the overall uncertainty of the $\left|\left\langle\sigma\tau \right\rangle\right|$ values. As can be seen from Tables~\ref{tab:qec} and \ref{tab:ft}, the precisions of the $ft$ and $\left|\left\langle\sigma\tau\right\rangle\right|$ values are still limited by the uncertainties in the half-lives and branching ratios.  

The $Q_\text{EC}$ value of $^{58}$Cu is important for the calibration of the $B(GT)$ values in $^{58}$Ni($^3$He,t)$^{58}$Cu charge-exchange reactions \cite{Fuj07}. The measured $Q_\text{EC}$ value, the half-life of $3.204(7)~\text{s}$ \cite{Fre65b} and an average branching ratio of $81.1(4)~\text{\%}$ for $^{58}$Cu (from the values of $80.8(7)~\text{\%}$ \cite{Per01}, $81.2(5)~\text{\%}$ \cite{Jan01} and $82(3)~\text{\%}$ \cite{Jon72}) yields $\log~ft=4.8701(24)$ with the calculator in Ref.~\cite{nndc}. The obtained Gamow-Teller strength is $B(GT)=0.08285(46)$ and the squared Gamow-Teller matrix element is $\left\langle \sigma\tau\right\rangle^2=0.05141(33)$. The values are little higher and more precise than previously (c.f. $B(GT)=0.0821(7)$ and $\left\langle \sigma\tau\right\rangle^2=0.0512(5)$ in Ref.~\cite{Per01}).

\begin{table}[!]
\caption{\label{tab:qec} The $Q_\text{EC}$ values determined in this work. The values have been measured directly except for $^{60}$Zn for 
which the adjusted mass excess value has been used.}
\begin{ruledtabular}
\begin{tabular}{llll}
Nuclide & $Q_\text{EC}$ (keV) & $Q_{EC, AME}$ \cite{AME} (keV) & JYFL-AME (keV) \\
\hline
$^{53}$Co & $8288.12(45)$ 		& $8300(18)$ & $-12(18)$ \\
$^{53}$Co$^m$ & $11462.2(12)$\footnotemark[1] & $11498(22)$ & $-36(22)$ \\
$^{55}$Ni & $8694.04(58)$  		& $8692(11)$ & $2(11)$ \\
$^{56}$Ni & $2132.76(46)$ 		& $2136(11)$ & $-3(11)$ \\
$^{57}$Cu & $8775.07(51)$ 		& $8772(16)$ & $3(16)$ \\ 
$^{58}$Cu & $8561.00(46)$ 		& $8565.6(14)$ & $-4.6(15)$\\
$^{59}$Zn & $9142.82(67)$ 		& $9097(40)$ & $46(40)$\\
$^{60}$Zn & $4171.4(18)$\footnotemark[2] & $4156(11)$ & $15(11)$\\
\end{tabular}
\end{ruledtabular}
\footnotetext[1]{$Q_\text{EC}$ value to the $^{53}$Fe ground state.}
\footnotetext[2]{Based on the measured mass excess value and the mass of $^{60}$Cu from Ref.~\cite{AME}.}
\end{table}

\begin{table}[!]
\caption{\label{tab:ft} The half-lives, electron capture probabilities ($P_{EC}$), branching ratios (BR), uncorrected $\log~ft$ values and experimental Gamow-Teller matrix elements $\left|\left\langle \sigma\tau \right\rangle\right|_{exp}$ for the mirror nuclei studied in this work. The average half-lives and branching ratios have been taken from Ref.~\cite{Sev08}. The calculator from \cite{nndc} was used for the $P_{EC}$ and $\log~ft$ values.}
\begin{ruledtabular}
\begin{tabular}{llllll}
Parent & $t_{1/2}$  & $P_{EC}$  & $BR$ &  $\log~ft$  & $\left|\left\langle \sigma\tau \right\rangle\right|_{exp}$ \\
nucleus & (ms) & (\%) & (\%)  &  \\
\hline
$^{53}$Co & 244.6(76) & 0.099(2) & 94.4(17) & 3.625(17) & 0.532(33)\\  
$^{55}$Ni & 203.3(37) &  0.103(1) & 100(10) & 3.62(5) &  0.54(10)\\
$^{57}$Cu & 196.44(68) & 0.103(1) & 89.9(8) & 3.670(5) & 0.441(11) \\
$^{59}$Zn & 181.9(18) & 0.107(1) & 94.03(77) & 3.706(6) & 0.360(14) \\
\end{tabular}
\end{ruledtabular}
\end{table}

\subsection{\label{sec:cde} Coulomb displacement energies}

If charge symmetry is assumed, the energy difference between the isobaric analog states (IAS) in mirror nuclei is only due to the Coulomb interaction and neutron-proton mass difference. If charge independence is assumed, the same is also true for isobaric triplets with $T=1$.  

Coulomb displacement energy (CDE) is the total binding energy difference between the isobaric analog states in the neighboring isobars determined as $CDE=Q_{EC}+\Delta _{n-H}$ where $\Delta _{n-H}=782.34660(55)~\text{keV}$ is the neutron-hydrogen mass difference. The Coulomb displacement energies follow a straight line when plotted as a function of $(Z-0.5)/A^{1/3}$ (see, \emph{e.g.}, Ref.~\cite{Ant97}) if a simple model for an evenly charged spherical nucleus is assumed. Deviations from the line reflect structural changes in the nuclei.

\begin{figure}[!]
\center{
\resizebox{0.45\textwidth}{!}{
\includegraphics{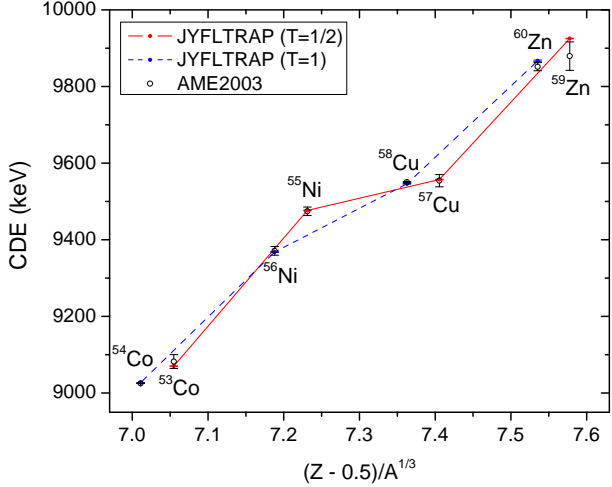}}
\caption{(Color online) Coulomb displacement energies for the $T=1/2$ doublets and $T=1$ triplets in Co, Ni, Cu and Zn isotopes from JYFLTRAP measurements and AME03 \cite{AME}. The JYFLTRAP $Q_\text{EC}$ values are from this work except the value for $^{54}$Mn which is from Ref.~\cite{Ero08}. For the $T=1$ states, excitation energies have been taken into account. For the $^{60}$Zn - $^{60}$Cu pair, a $0^+,~T=1$ state is not known and therefore a $2^+,~T=1$ has been used.}
\label{fig:cde}       
}
\end{figure}

Coulomb displacement energies from JYFLTRAP for $T=1/2$ mirror and $T=1$ isobaric analog states of cobalt, nickel, copper, and zinc nuclides are plotted in Fig.~\ref{fig:cde}. As can be seen from Fig.~\ref{fig:cde}, the CDE values do not follow a straight line as a function of $(Z-0.5)/A^{1/3}$. This can be partly explained by different spins in the $T=1/2$ states: the ground state spin changes from $7/2^-,~T=1/2$ to $3/2^-,~T=1/2$ at $^{57}$Cu. As the protons in the $p$ orbits have a larger radius than the protons in the $f$ orbits, the Coulomb repulsion in $^{53}$Co and $^{55}$Ni filling the $1f_{7/2}$ proton shells is stronger than in $^{57}$Cu and $^{59}$Zn filling the $2p_{3/2}$ shells. Compared to the AME03 \cite{AME} values, the precision of the CDE values has now been improved considerably and deviations have been found for $^{58}$Cu, $^{59}$Zn, and $^{60}$Zn. The trend is a little smoother in the $T=1$ states. There, it should be noted that the $0^+,~T=1$ state is not always the ground state. The lowest $T=1,~0^+$ state lies at $1450.68(4)~\text{keV}$ in $^{56}$Co, at $7903.7(10)~\text{keV}$ in $^{56}$Ni, and at $202.6(3)~\text{keV}$ in $^{58}$Cu. For $^{60}$Zn, the $0^+,~T=1$ level is not known but the level at $4913.1(9)~\text{keV}$ is probably a $T=1$ analogue state of the $^{60}$Cu $2^+$ ground state \cite{NDS100} which has been adopted in Fig.~\ref{fig:cde}. For the other nuclides in Fig.~\ref{fig:cde}, the  $0^+, T=1$ state is the ground state.

\subsection{\label{sec:Sp} Proton-capture $Q$ values for the $rp$ process}

Proton separation energies $S_p$ (or proton-capture $Q$ values) can be measured directly in a similar way as the $Q_\text{EC}$ values with a Penning trap. From the measured frequency ratio $r$ between a nuclide $(Z,A)$ with a mass $m_m$ and the reference $(Z-1,A-1)$ with a mass $m_d$, a proton separation energy is obtained as:

\begin{equation}
\label{eq:sp}
\begin{split}
S_p &= (-m_m+m_d+m_H)c^2\\
	&= \left[(1-r)(m_d-m_e)+m_H)\right]c^2\\
\end{split}
\end{equation}

where $m_H$ is the mass of a hydrogen atom.

With this method, $S_p$ values for $^{56}$Ni, $^{57}$Cu, $^{59}$Zn, and $^{60}$Zn were measured directly (see Table~\ref{tab:sp}). The $S_p$ values for $^{53}$Co, $^{55}$Ni, and $^{58}$Cu were also improved with the new mass values of this work. The biggest differences to the AME03 values occur at $^{59}$Zn and $^{60}$Zn which are now less proton-bound. The $S_p$ value of $^{58}$Cu differs slightly from the AME03 value.

\begin{table}[!]
\caption{\label{tab:sp} The $S_p$ values determined in this work and comparison to the literature values \cite{AME}. The $S_p$ values were 
measured directly with respect to the proton-decay daughters for $^{56}$Ni, $^{57}$Cu, $^{59}$Zn and $^{60}$Zn. For the others, mass 
excess values determined in this work and literature values from Ref.~\cite{AME} have been used.}
\begin{ruledtabular}
\begin{tabular}{llll}
Nuclide & $S_p$ (keV) & $S_{p, AME}$ \cite{AME} (keV) & JYFL-AME (keV)\\
\hline
$^{53}$Co\footnotemark[1] & 1615(7) & 1602(19)  & 13(20) \\
$^{54}$Ni\footnotemark[2] & 3842(50) & 3855(50) & -13(70) \\
$^{55}$Ni\footnotemark[3] & 4615.7(11) & 4617(11) & -1(11) \\
$^{56}$Ni & 7165.84(45) & 7165(11) & 1(11) \\
$^{56}$Cu\footnotemark[2] & 555(140) & 554(140) & 1(200) \\
$^{57}$Cu & 689.69(51) & 695(19) & -5(19) \\ 
$^{58}$Cu\footnotemark[1] & 2872.55(76) & 2869.1(24)  & 3.5(25)\\
$^{58}$Zn\footnotemark[2] & 2279(50) & 2277(50) & 2(70) \\
$^{59}$Zn & 2837.21(91)\footnotemark[4] & 2890(40) & -53(40) \\
$^{60}$Zn  & 5104.93(51)  & 5120(11)  & -15(11) \\
$^{60}$Ga\footnotemark[2] & 73(110) & 26(120)  & 50(160)\\
$^{61}$Ga\footnotemark[2] & 207(50) & 192(50)  & 15(80)\\
\end{tabular}
\end{ruledtabular}
\footnotetext[1]{The value of the proton-decay daughter of the nuclide has been taken from Ref.~\cite{AME}.}
\footnotetext[2]{The mass excess value for the proton-decay daughter is from this work whereas the mass excess value for the nuclide is 
from Ref.~\cite{AME}.}
\footnotetext[3]{The mass excess value for the proton-decay daughter $^{54}$Co was taken from Ref.~\cite{Ero08}.}
\footnotetext[4]{$S_p$ calculated from the adjusted mass values given in Table~\ref{tab:MEadj}. The directly measured value, $S_p=2836.9(12)~\text{keV}$, is less precise due to a large uncertainty in $\delta r$.}
\end{table}

\begin{figure}[!]
\centering
\includegraphics[width=0.45\textwidth,clip]{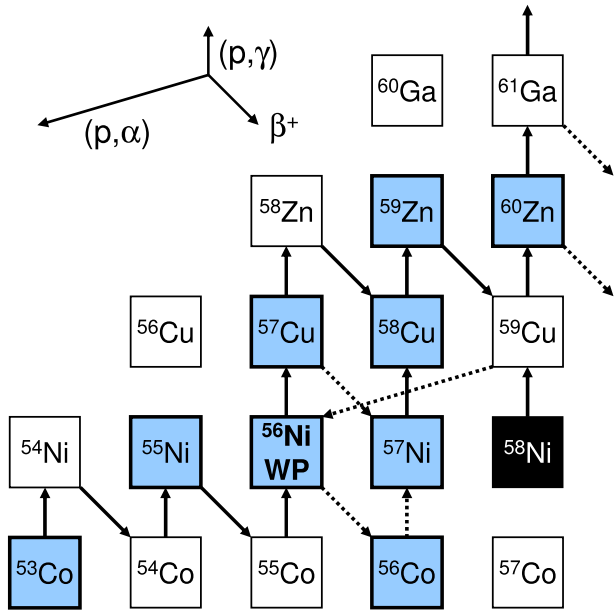}
\caption{(Color online) The $rp$-process path for steady-state burning conditions according to Ref.~\cite{Sch01}. Shown are the reaction flows of more than $10\%$ (solid line) and of $1\%-10\%$ (dashed line) of the reaction flow through the $3\alpha$ reaction. All of the measured nuclides (highlighted) lie at the $rp$-process path flowing through the waiting-point (WP) nucleus $^{56}$Ni.}
\label{fig:rp_path}      
\end{figure}

In this work, we have improved the precisions of the $Q$ values for the proton captures as well as the $Q_\text{EC}$ values for the nuclides shown in Fig.~\ref{fig:rp_path}. This helps to do more reliable astrophysical calculations for the $rp$ process. The $rp$-process path beyond the waiting-point nucleus $^{56}$Ni is of special interest and is mainly determined by the $^{56}$Ni(p,$\gamma$)$^{57}$Cu proton capture rate at lower temperatures (below $\approx 1 ~\text{GK}$) and the $\beta^+$-decay rate of $^{58}$Zn at higher temperatures. The reaction rate for $^{56}$Ni(p,$\gamma$)$^{57}$Cu has been calculated in detail in Ref.~\cite{For01}. 

\begin{table*}[!]
\caption{\label{tab:rprate} Resonance parameters for the reaction $^{56}$Ni(p,$\gamma$)$^{57}$Cu. The resonance energy $E_r$ has been determined with the new JYFLTRAP $S_p$ value for $^{57}$Cu. The values of $\Gamma_p$ have been scaled from Ref.~\cite{For01} with new $E_r$ values and the $\Gamma_\gamma$ values are directly from \cite{For01}.}
\begin{ruledtabular}
\begin{tabular}{llllll}
$E_x$ (keV) \footnotemark[1] &  $J_f$ \footnotemark[1]  & $E_r$ (MeV) & $\Gamma_p$ (eV) \footnotemark[2] & $\Gamma_\gamma$ (eV)\footnotemark[3] & $\omega\gamma$ (eV)\\
\hline
1028(4) & $5/2^-$ & 0.338(4) & $7.94\times 10^{-12}$ & $3.55\times 10^{-4}$ & $2.38\times 10^{-11}$ \\
1106(4) & $1/2^-$ & 0.416(4) & $2.26\times 10^{-7}$ & $4.23\times 10^{-3}$ & $2.26\times 10^{-7}$ \\
2398(10) & $5/2^-$ & 1.708(10) & $9.43\times 10^{-2}$ & $1.37\times 10^{-2}$ & $3.59\times 10^{-2}$ \\
2520(25) & $7/2^-$ & 1.830(25) & $7.29\times 10^{-2}$ & $8.38\times 10^{-3}$ & $3.01\times 10^{-2}$ \\
\end{tabular}
\end{ruledtabular}
\footnotetext[1]{From Ref.~\cite{Bha98}.}
\footnotetext[2]{Scaled from the values in Ref.~\cite{For01}}
\footnotetext[3]{From Ref.~\cite{For01}.}
\end{table*}

The $Q$ value for the reaction $^{56}$Ni(p,$\gamma$)$^{57}$Cu has now been improved from $695(19)~\text{keV}$ to $689.69(51)~\text{keV}$. A new reaction rate can be estimated with the new resonance energies $E_r=E_x-S_p$, where $E_x$ is the excitation energy of the final state in $^{57}$Cu and $S_p$ is the proton separation energy for $^{57}$Cu. The astrophysical reaction rate for resonant captures to states with resonance energies $E_i$ and resonance strengths $\omega\gamma_i$ (both in MeV) is obtained by:

\begin{equation}
\label{eq:rate}
\begin{split}
N_A\left\langle \sigma v\right\rangle_r &= 1.54\times10^{11}(\mu T_9)^{-3/2}\sum_i(\omega \gamma)_i \times\\
	& \exp(-11.605E_i/T_9)~\text{cm}^3\text{mol}^{-1}\text{s}^{-1}~.\\
\end{split}
\end{equation}

The resonance strength $\omega\gamma$ for an isolated resonance in a $(p,\gamma)$ reaction is given by:
\begin{equation}
\label{eq:strength}
\omega\gamma = \frac{2J+1}{2(2J_t+1)}\frac{\Gamma_p\Gamma_\gamma}{\Gamma_\text{tot}}\\
\end{equation}

where $J$ and $J_t$ are the spins of the resonance state and the target nucleus ($^{56}$Ni) and the total width $\Gamma_\text{tot}$ is the sum of the proton width $\Gamma_p$ and the gamma width $\Gamma_\gamma$. The proton widths have now been scaled from Ref.~\cite{For01} by using the relation:

\begin{equation}
\label{eq:scale}
\Gamma_p \propto \exp\left(-31.29Z_1Z_2\sqrt{\frac{\mu}{E_r}}\right)\\
\end{equation}
 
where $Z_1$ and $Z_2$ are the proton numbers of the incoming particles, $\mu$ is the reduced mass in u and $E_r$ is the center-of-mass resonance energy in keV \cite{Rol88}. The resonance parameters are summarized in Table~\ref{tab:rprate}. The non-resonant reaction rate has been taken from \cite{For01} and scaled with the new value for the reduced mass. 

With the new $Q$ value, a factor of four in the uncertainty of the reaction rate at temperatures around $1~\text{GK}$ shown in Ref.~\cite{For01} is removed and the new rate is a little higher than calculated with the old $Q$ value (see Fig.~\ref{fig:rate}). The new $Q$ value supports the conclusions of Ref.~\cite{For01} that the lifetime of $^{56}$Ni against proton capture is much shorter than in the previous works. This reduces the minimum temperature required for the $rp$ process to proceed beyond $^{56}$Ni. In fact, with the rates of Ref.~\cite{For01}, this temperature threshold coincides with the temperature for the break out of the hot CNO cycles. 

\begin{figure}[!]
\centering
\includegraphics[width=0.45\textwidth,clip]{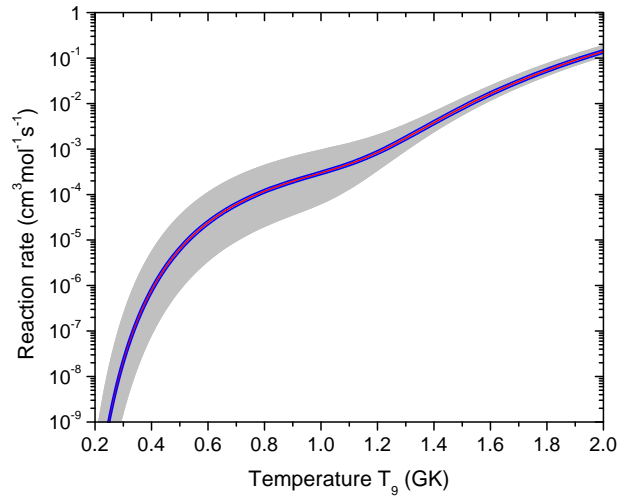}
\caption{(Color online) Total reaction rate for $^{56}$Ni(p,$\gamma$)$^{57}$Cu \label{fig:rate}. The grey-shaded area shows the calculated reaction rate with the old resonance energies including the uncertainties coming from the reduced mass $\mu$, $S_p$ for $^{57}$Cu, and resonance energies $E_i$. The red curve has been plotted with the new JYFLTRAP values for the reduced mass and $S_p$ of $^{57}$Cu and the blue curves show the error band including the uncertainties in $\mu$, $S_p$ and $E_i$. The precise $S_p$ value reduces significantly the uncertainties of the calculated reaction rate. }
\label{fig:rate}     
\end{figure}

\section{\label{sec:summary}Summary and conclusions}

In conclusion, atomic masses in the vicinity of the doubly-magic $^{56}$Ni nucleus have been measured with the JYFLTRAP Penning trap mass spectrometer. Frequency ratios measured between 13 nuclides close to $A=56$ formed an overdetermined network for which a least-squares minimization has been done. The adjusted mass values have improved the precisions of the AME03 mass values remarkably. The most surprising deviations to the AME03 have been found at $A=58$. The AME03 value for $^{58}$Cu based on $(p,n)$ threshold energy measurements deviates from the value obtained in this work by $2.2\sigma$. For $^{58}$Ni, a $1\sigma$ deviation to the AME03 value has been found but the value agrees almost perfectly with the older $(n,\gamma)$ results for $^{58}$Ni \cite{Wil75,Ish77}. In addition, the mass values obtained for $^{55}$Co, $^{59}$Zn, and $^{60}$Zn deviate from the AME03 values by $1.3-1.4\sigma$. 

The excitation energy of the proton-emitting $19/2^-$ isomeric state in $^{53}$Co has been improved and the decay scheme of $^{53}$Co revised. The $Q_\text{EC}$ values for $T=1/2$ mirror transitions of $^{53}$Co, $^{55}$Ni, $^{57}$Cu, and $^{59}$Zn have been measured directly. These values are useful for precise weak interaction information and a possible derivation of $|V_{ud}|$ in future when either of the beta or neutrino asymmetry parameter or $\beta-\nu$ angular correlation coefficient have been measured precisely enough. Coulomb displacement energies between isobaric analog states have been determined from these mirror $Q_\text{EC}$ values and from the $Q_\text{EC}$ values of $^{56}$Ni and $^{58}$Cu. The new $Q_\text{EC}$ value for the $^{58}$Cu ground-state beta decay allowed a revised $B(GT)$ value for this transition used as a calibrant in $^{58}\text{Ni}(^3\text{He,t})$ charge-exchange reaction studies. All measured nuclides lie at the path of the astrophysical $rp$ process for which the improved $S_p$ and $Q_\text{EC}$ values are important. In particular, we have directly measured the $Q$ value for the proton capture on $^{56}$Ni which removes the large uncertainties in the corresponding reaction rate at lower temperatures. The new result supports the conclusions of earlier works that the $rp$ process can proceed beyond $^{56}$Ni.

\begin{acknowledgments}
This work has been supported by the EU 6th Framework programme ``Integrating Infrastructure Initiative - Transnational Access", Contract Number: 506065 (EURONS) and by the Academy of Finland under the Finnish Centre of Excellence Programme 2006-2011 (Nuclear and Accelerator Based Physics Programme at JYFL). A.K. acknowledges the support from the Academy of Finland under the project 127301. 
\end{acknowledgments}

\bibliography{aps}

\begin{thebibliography}{a56}
\bibitem{Sch98} H. Schatz \emph{et al.}, Phys. Rep. \textbf{294}, 167 (1998).
\bibitem{Cru92} M.T.F. da Cruz \emph{et al.}, Phys. Rev. C \textbf{46}, 1132 (1992).
\bibitem{Wal81} R. K. Wallace and S. E. Woosley, Astrophys. J. Suppl. Series \textbf{45}, 389 (1981).
\bibitem{Sch01} H. Schatz \emph{et al.}, Phys. Rev. Lett. \textbf{86}, 3471 (2001).
\bibitem{Elo09a} V.-V. Elomaa, G. K. Vorobjev, A. Kankainen \emph{et al.}, Phys. Rev. Lett. \textbf{102}, 252501 (2009).
\bibitem{Cer70} J. Cerny \emph{et al.}, Phys. Lett. \textbf{33B}, 284 (1970).
\bibitem{Cer72} J. Cerny \emph{et al.}, Nucl. Phys. \textbf{A188}, 666 (1972).
\bibitem{Sev08} N. Severijns, M. Tandecki, T. Phalet, and I. S. Towner, Phys. Rev. C \textbf{78}, 055501 (2008). 
\bibitem{Nav08} O. Naviliat-Cuncic and N. Severijns, Phys. Rev.Lett. \textbf{102}, 142302 (2009).
\bibitem{Ays01} J. \"Ayst\"o, Nucl. Phys. \textbf{A693}, 477 (2001).
\bibitem{Elo08}  V.-V. Elomaa \emph{et al.}, Eur. Phys. J. A \textbf{40}, 1 (2009).
\bibitem{nubase} G. Audi, O. Bersillon, J. Blachot and A.H. Wapstra, Nucl. Phys. \textbf{A729}, 3 (2003).
\bibitem{Elo09} V.-V. Elomaa \emph{et al.}, Nucl. Instrum. and Methods in Phys. Res., Sec. A \textbf{612}, 97 (2009).
\bibitem{Hui02} J. Huikari \emph{et al.}, Nucl. Instrum. and Methods in Phys. Res., Sec. B \textbf{222}, 632 (2004). 
\bibitem{Kan05} A. Kankainen \emph{et al.}, Eur. Phys. J. A \textbf{29}, 271 (2006).
\bibitem{Web08} C. Weber \emph{et al.}, Phys. Rev. C \textbf{78}, 054310 (2008).
\bibitem{Rah08} S. Rahaman \emph{et al.}, Phys. Lett. B \textbf{662}, 111 (2008).
\bibitem{Nie01} A. Nieminen \emph{et al.}, Nucl. Instrum. Methods Phys. Res., Sec. A \textbf{469}, 244 (2001).
\bibitem{Kol04} V.S. Kolhinen \emph{et al.}, Nucl. Instrum. Methods Phys. Res., Sec. A \textbf{528}, 776 (2004).
\bibitem{Sav91} G. Savard \emph{et al.}, Phys. Lett. A \textbf{158}, 247 (1991).
\bibitem{Gra80} G. Gr\"aff, H. Kalinowsky, and J. Traut, Z. Phys. A \textbf{297}, 35 (1980).
\bibitem{Kon95} M. K\"onig \emph{et al.}, Int. J. Mass. Spectrom. Ion Process. \textbf{142}, 95 (1995).
\bibitem{Gab09} G. Gabrielse, Phys. Rev. Lett. \textbf{102}, 172501 (2009).
\bibitem{Geo07} S. George \emph{et al.}, Int. J. Mass Spectrom. \textbf{264}, 110 (2007).
\bibitem{Kre07} M. Kretzschmar, Int. J. Mass Spectrom. \textbf{264}, 122 (2007).
\bibitem{Ero09} T. Eronen \emph{et al.}, Phys. Rev. Lett. \textbf{103}, 252501 (2009).
\bibitem{Kel03} A. Kellerbauer \emph{et al.}, Eur. Phys. J. D \textbf{22}, 53 (2003).
\bibitem{Bir32} R. Birge, Phys. Rev. \textbf{40}, 207 (1932).
\bibitem{AME} G. Audi, A.H. Wapstra and C. Thibault, Nucl. Phys. A \textbf{729}, 337 (2003).
\bibitem{Aud86} G. Audi, W.G. Davies and G.E. Lee-Whiting, Nucl. Instrum. and Methods in Phys. Res., Sec. A \textbf{249}, 443 (1986).
\bibitem{Muk08} M. Mukherjee \emph{et al.}, Eur. Phys. J. A \textbf{35}, 31 (2008).
\bibitem{Mue75} D. Mueller \emph{et al.}, Phys. Rev. C \textbf{12}, 51 (1975).
\bibitem{Wil03} S. Williams \emph{et al.}, Phys. Rev. C \textbf{68}, 011301(R) (2003).
\bibitem{Ben07} M.A. Bentley and S.M. Lenzi, Progr. in Part. and Nucl. Phys. \textbf{59}, 497 (2007).
\bibitem{Pro72} I.D. Proctor et al., Phys. Rev. Lett. \textbf{29}, 434 (1972).
\bibitem{Mue77} D. Mueller, E. Kashy, and W. Benenson, Phys. Rev. C \textbf{15}, 1282 (1977).
\bibitem{Ays84} J. \"Ayst\"o \emph{et al.}, Phys. Lett. \textbf{138B}, 369 (1984).
\bibitem{Hoo65} C.G. Hoot, M. Kondo and M.E. Rickey, Nucl. Phys. \textbf{71}, 449 (1965).
\bibitem{Mil67} R.G. Miller and R.W. Kavanagh, Nucl. Phys. \textbf{A94}, 261 (1967).
\bibitem{Joh08} E.K. Johansson \emph{et al.}, Phys. Rev. C \textbf{77}, 064316 (2008).
\bibitem{Shi84} T. Shinozuka \emph{et al.}, Phys. Rev. C \textbf{30}, 2111 (1984).
\bibitem{She85} B. Sherrill \emph{et al.}, Phys. Rev. C \textbf{31}, 875 (1985).
\bibitem{Gag86} C.A. Gagliardi, D.R. Semon, R.E. Tribble, and L.A. Van Ausdeln, Phys. Rev. C \textbf{34}, 1663 (1986).
\bibitem{Sti87} E. Stiliaris \emph{et al.}, Z. Phys. A \textbf{326}, 139 (1987).
\bibitem{Fre65} J.M. Freeman \emph{et al.}, Nucl. Phys. \textbf{65}, 113 (1965).
\bibitem{Bon66} B.E. Bonner \emph{et al.}, Nucl. Phys. \textbf{86}, 187 (1966).
\bibitem{Ove69} J.C. Overley, P.D. Parker and D.A. Bromley, Nucl. Instrum. and Methods in Phys. Res., Sec. A \textbf{68}, 61 (1969).
\bibitem{Rud02} D. Rudolph \emph{et al.}, Eur. Phys. J. A \textbf{14}, 137 (2002).
\bibitem{Fre76} J.M. Freeman, Nucl. Instrum. Methods \textbf{134}, 153 (1976).
\bibitem{Mar66} J.B. Marion, Rev. Mod. Phys. \textbf{38}, 660 (1966).
\bibitem{Ero08} T. Eronen \emph{et al.}, Phys. Rev. Lett. \textbf{100}, 132502 (2008).
\bibitem{Ara81} Y. Arai \emph{et al.}, Phys. Lett. \textbf{104B}, 186 (1981).
\bibitem{She83} B. Sherrill \emph{et al.}, Phys. Rev. C \textbf{28}, 1712 (1983).
\bibitem{Gre72} M.B. Greenfield \emph{et al.}, Phys. Rev. C \textbf{6}, 1756 (1972).
\bibitem{Kaw86} H. Kawakami \emph{et al.}, J. Phys. Soc. Jpn. \textbf{55}, 3014 (1986).  
\bibitem{Mar72} D.J. Martin \emph{et al.}, Nucl. Phys. \textbf{A187}, 337 (1972).
\bibitem{Erl77} B. Erlandsson and J. Lyttkens, Z. Phys. A \textbf{280}, 79 (1977).
\bibitem{Han80} R. H\"anninen and G.U. Din, Phys. Scripta \textbf{22}, 439 (1980).
\bibitem{Gos73} J.D. Goss, C.P. Browne and A.A. Rollefson, Phys. Rev. Lett. \textbf{30}, 1255 (1973).
\bibitem{Jol74} P.L. Jolivette \emph{et al.}, Phys. Rev. C \textbf{10}, 2449 (1974).
\bibitem{Pet68} H. Pettersson, O. Bergman and C. Bergman, Arkiv f\"or Fysik \textbf{29}, 423 (1968).
\bibitem{Wil75} W.M. Wilson, G.E. Thomas and H.E. Jackson, Phys. Rev. C \textbf{11}, 1477 (1975).
\bibitem{Ish77} A.F.M. Ishaq \emph{et al.}, Z. Phys. A \textbf{281}, 365 (1977).
\bibitem{Guo92} Z. Guo \emph{et al.}, Nucl. Phys. \textbf{A540} (1992) 117.
\bibitem{Spi68} P. Spilling \emph{et al.}, Nucl. Phys. \textbf{A113}, 395 (1968).
\bibitem{Alb76} D.E. Alburger,  Nucl. Instrum. and Methods in Phys. Res. \textbf{136}, 323 (1976).
\bibitem{Ste78} M.L. Stelts and R.E. Chrien, Nucl. Instrum. and Methods in Phys. Res. \textbf{155}, 253 (1978).
\bibitem{Isl80} M.A. Islam \emph{et al.}, Can. J. Phys. A \textbf{58}, 168 (1980).
\bibitem{Ven80} R. Vennink \emph{et al.},  Nucl. Phys. \textbf{A344}, 421 (1980).
\bibitem{Nan76} H. Nann \emph{et al.}, Phys. Rev. C \textbf{14}, 2338 (1976).
\bibitem{Gue07} C. Gu\'enaut \emph{et al.}, Phys. Rev. C \textbf{75}, 044303 (2007).
\bibitem{Har93} A. Harder \emph{et al.}, Z. Phys. A \textbf{345}, 143 (1993).
\bibitem{Ram02} S. Raman private comm. from Ref.~\cite{AME}.
\bibitem{Fir03} R.B. Firestone \emph{et al.} IAEA-Tecdoc 5 (2003). 
\bibitem{Bon63} R.O. Bondelid and J.W. Butler, Phys. Rev. \textbf{130}, 1078 (1963).
\bibitem{Fod70} I. Fodor, I. Szentp\'etery and J. Sz\"ucs, Phys. Lett. \textbf{32B}, 689 (1970).
\bibitem{Kla75} H.V. Klapdor \emph{et al.}, Nucl. Phys. \textbf{A245}, 133 (1975).
\bibitem{NDS87} H. Junde, Nucl. Data Sheets \textbf{87}, 507 (1999).
\bibitem{nndc} http://www.nndc.bnl.gov/logft/
\bibitem{Tow10} I.S. Towner and J.C. Hardy, Rep. Prog. Phys. \textbf{73}, 046301 (2010).
\bibitem{PDG} C. Amsler \emph{et al.} (Particle Data Group), Phys. Lett. B \textbf{667}, 78 (2008).
\bibitem{Fuj07} H. Fujita, Y. Fujita \emph{et al.}, Phys. Rev. C \textbf{75}, 034310 (2007).
\bibitem{Fre65b} J.M. Freeman \emph{et al.}, Nucl. Phys. \textbf{69}, 433 (1965).
\bibitem{Per01} K. Per\"aj\"arvi \emph{et al.}, Nucl. Phys. \textbf{A696}, 233 (2001).
\bibitem{Jan01} Z. Janas \emph{et al.}, Eur. Phys. J. A \textbf{12}, 143 (2001).
\bibitem{Jon72} H.W. Jongsma \emph{et al.}, Nucl. Phys. \textbf{A179}, 554 (1972).
\bibitem{Ant97} M.S. Antony, A. Pape, and J. Britz, At. Data and Nucl. Data Tabl. \textbf{66}, 1 (1997). 
\bibitem{NDS100} J.K. Tuli, Nucl. Data Sheets \textbf{100}, 347 (2003).
\bibitem{For01} O. Forstner, H. Herndl, H. Oberhummer, H. Schatz, and B.A. Brown, Phys. Rev. C \textbf{64}, 045801 (2001).
\bibitem{Rol88} C. E. Rolfs and W. S. Rodney, in Cauldrons in the Cosmos, edited by D. N. Schramm (The University of Chicago Press, Chicago, 1988).
\bibitem{Bha98} M.R. Bhat, Nucl. Data Sheets \textbf{85}, 415 (1998).
\end{thebibliography}
\end{document}